\documentclass{aastex63}

\usepackage{subfigure}

\submitjournal{ApJS}

\shorttitle{Radio and HE/VHE sources in Cygnus}
\shortauthors{Benaglia et al.}

\begin{document}

\title{Radio counterparts of gamma-ray sources in the Cygnus region}

\correspondingauthor{Paula Benaglia}
\email{paula@iar-conicet.gov.ar, pben.radio@gmail.com}

\author[0000-0002-6683-3721]{Paula Benaglia}
\affiliation{Instituto Argentino de Radioastronom\'{\i}a, CONICET-CICPBA-UNLP, CC5 (1897) Villa Elisa, Prov. de Buenos Aires, Argentina}


\author{C. H. Ishwara-Chandra}
\affiliation{National Centre for Radio Astrophysics, Tata Institute of Fundamental Research, Pune University Campus, Pune, 411007, India}

\author[0000-0002-1566-9044]{Josep M. Paredes}
\affiliation{Departament de Física Quàntica i Astrofísica, Institut de Ciències del Cosmos, Universitat de Barcelona, IEEC-UB, Martí i Franquès 1, 08028, Barcelona, Spain}

\author[0000-0002-5880-2730]{Huib T. Intema}
\affiliation{International Centre for Radio Astronomy Research, Curtin University, Bentley, WA 6102, Australia}
\affiliation{Leiden Observatory, Leiden University, Niels Bohrweg 2, 2333 CA Leiden, the Netherlands}

\author{Marcelo E. Colazo}
\affiliation{Comisi\'{o}n Nacional de Actividades Espaciales, Paseo Col\'{o}n 751 (1063) CABA, Argentina}

\author{Natacha L. Isequilla}
\affiliation{Instituto de Astronomía y Física del Espacio, CONICET-UBA, CC 67, Suc. 28, 1428, Buenos Aires, Argentina}

\begin{abstract}
The view of the gamma-ray universe is being continuously expanded by space high energy (HE) and ground based very-high energy (VHE) observatories. Yet, the angular resolution limitation still precludes a straightforward identification of these gamma-ray emitting sources. Radio observations are an effective tool for searching their possible counterparts at lower energies because the same population of relativistic electrons responsible for radio emission can also produce HE/VHE emission via inverse-Compton scattering. The Cygnus region is crowded by many gamma-ray sources, most of them remaining unidentified. In order to find possible counterparts to unidentified gamma-ray sources, we carried out a deep survey of the Cygnus region using the Giant Metrewave Radio Telescope at 610 MHz and 325 MHz. We did a detailed search for counterparts in the error circle of HE/VHE sources. We report 36 radio sources found in the error ellipse of 15 HE sources, and 11 in those of VHE sources. Eight sources have very steep radio spectral index $\alpha <-1.5$, which are most likely to be pulsars and will be followed up for periodicity search. Such a significant number of pulsar candidates within the error circle of HE/VHE sources prompts fresh look at the energetics and efficacy of pulsars and pulsar wind nebulae in this context. 
\end{abstract}


\section{Introduction}\label{sec:intro}

From the first all-sky gamma-ray surveys and associated catalogs, a large fraction of the discovered sources lacked identification with sources detected along the rest of the electromagnetic spectrum \citep[e.g., ][]{swanenburg1981}.  Since the early days of observations of the gamma-ray sky, emission at these high energies was detected by known and of wide range of sources like blazars, radio galaxies and quasars of the extra-galactic origin and pulsars, supernova remnants and micro-quasars in our Galaxy.

At the GeV energy range, the Energetic Gamma Ray Experiment Telescope aboard the Compton Gamma Ray Observatory, provided a catalog of 271 sources, 170 of which were so-called unidentified gamma-ray sources or  UNIDS \citep{hartman1999}, with a position accuracy close to half of a degree. 
In the last decade, the Fermi Large Area Telescope relieved the sky with higher sensitivity and angular resolution ($\sim$0.1 deg). Its fourth Catalog \citep[4FGL, ][]{abdollahi2020} compiles information of five thousand sources above 4$\sigma$, and $\sim$1300 remain with unknown nature. The presence of more than  one - sometimes many - lower energy sources inside the position error of the { high energy (HE: 0.1 GeV $<$ E $<$ 100 GeV)} sources makes it difficult for  the accurate identification of the specific object producing gamma rays. 

 At very-high energies (VHE: above 100 GeV) a similar percentage of about ~22\% of unassociated sources is found for VHE sources detected by Cherenkov telescopes. Their distribution is strongly peaked towards the Galactic Plane, thus suggesting an important contribution from Milky Way emitters in this population. Undiscovered pulsars, pulsar wind nebulae or supernova remnants are likely among them. {Many studies to find lower energy counterpart’s candidates of selected HE/VHE sources have been undertaken during the last years by several authors. As an example, in the field of the unidentified VHE source HESS\, J1858+020 were found a supernovae remnant interacting with nearby molecular clouds and a few non-thermal sources \citep{paredes2014}. In the case of HE unidentified sources, we can mention 3FGL\, J0133.3+5930, where a galactic and an extragalactic objects were proposed as possible counterpart  \citep{marti2017}. }

The most energetic phenomena of the Universe give rise to gamma rays: non thermal sources with relativistic particles that interact with  fields, matter and radiation. Likewise, the combined ingredients { can} generate radiation at low radio frequencies, via the synchrotron mechanism. This radio radiation can be measured using radio interferometers, with  arcsecond resolution.  This superior resolution in the radio band as compared to the HE/VHE facilities will help to precisely localise possible counterparts. Moreover, multi-band observations provide  hints on the radiation mechanisms and then on  the kind of  object that generate the emission  measured. As example, \cite{frail2018} describe, in particular, a method to identify pulsars that can be physically related with unidentified Fermi sources. 
In short, though the fraction of the counterparts of gamma-ray sources have improved, there are still significant number of sources in the gamma-ray sky which do not have counterpart at any other branch of the electromagnetic spectrum \citep{abdollahi2020}.  

The purpose of this work was to collect evidence and contribute to the identification of discrete HE/VHE sources, by means of radio images at very high angular resolution and very low noise.
We focused on a group of such sources detected by the Fermi LAT instrument, and complement it with other sources at TeV energies. We surveyed a highly populated area of the northern sky nearby the Galactic plane at two bands centred at decimetre wavelengths, using the Giant  Metrewave Radio Telescope (GMRT). This work is part of a series of studies of the same celestial area in the Cygnus region, mainly the catalog introduced in \cite{benaglia2020b}. The latter lists radio sources at 610 MHz and 325 MHz bands, for flux densities greater or equal  to 7$\sigma$, where $\sigma$ is the local rms; see also \cite{ishwar2019,benaglia2020a,isequilla2020}.
Here we have studied sources above 3$\sigma$.
Section 2 describes the observed region including the  high-energy sources detected. Section 3 summarizes the data reduction process. In Section 4 we report the results. 
Section 5 explains the search for counterparts of the radio sources discovered here. A  discussion is presented in  Section 6, and we conclude in a last Section.

\section{The Cygnus region and the gamma-ray sources} \label{sec:region-sources}

The region under study corresponds to the central part of the Cygnus constellation  which is relatively nearby and hosts nine OB associations and various rich clusters \citep{uyaniker2001,mahy2013}, see also \cite{reipurthcyg} for a review. The region also has many unindentified VHE sources. The observed area covers $\sim$20 sq deg, centred at $RA, Dec$(J2000) = 20:25:30,~42:00:00. It encompasses the stellar association of Cyg\,OB2, and part of Cyg\,OB8 and OB9, extremely rich in OB stars and with traces of recent star formation 
\citep[see Figure\,2 by][]{benaglia2020b}.

\begin{figure*}[ht!]
\plotone{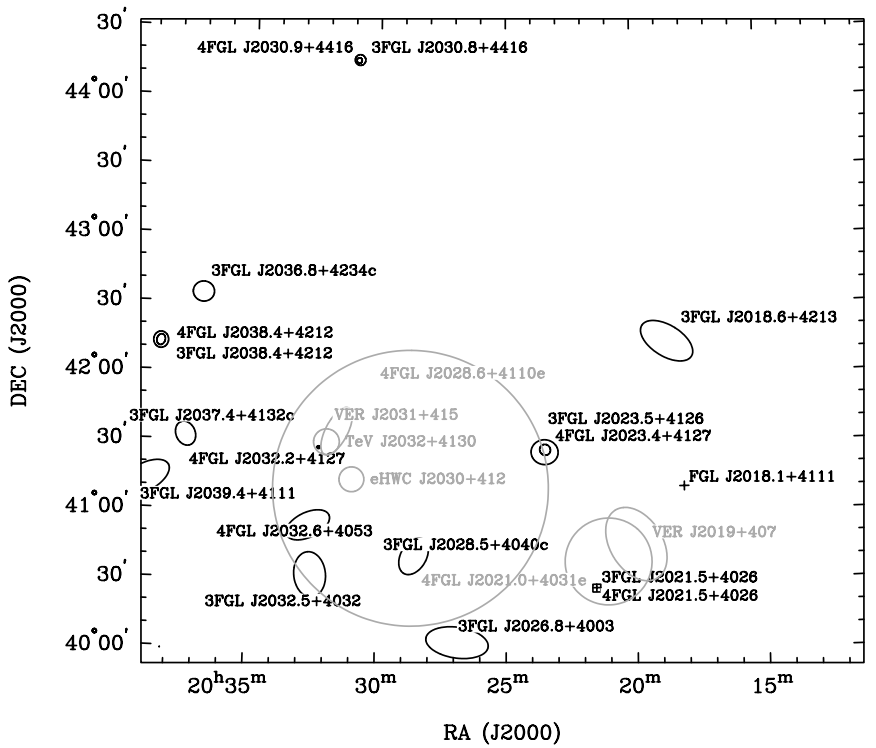}
\caption{Layout  of the HE and VHE sources in the observed region. Discrete GeV (Fermi) sources in black lines; extended GeV and  TeV sources in  light grey lines.
}
\label{fig:observedregion}
\end{figure*}

The Fermi LAT collaboration produced four all-purpose all-sky catalogs, namely First, Second, Third and Four Fermi Gamma-ray LAT (1FGL, 2FGL,  3FGL, 4FGL)  catalogs, after the  analysis of 0.9, 2, 4 and 8 years of data, respectively \citep[see ][and references therein]{abdollahi2020}. Besides different data collection periods and slightly different energy ranges (enlarged with date), the analysis software package and the  model of the  underlying diffuse emission to be subtracted were improved over time.  The 4FGL paper covered the range of  0.05~GeV to 1~TeV and detected 5064 point sources above 4$\sigma$; the authors reported that for 1336 they did not find plausible counterparts. The 4FGL catalog has a second release, the LAT 10-year Source Catalogue Data Release 2 (4FGL-DR2)\footnote{https://fermi.gsfc.nasa.gov/ssc/data/access/lat/10yr\_catalog/}. 

We selected the point sources of the 3FGL and 4FGL catalogs that lie in the Cygnus region observed at radio bands. There is one source only  present at the 4FGL catalog, five sources with 4FGL and 3FGL entries, and seven sources with only 3FGL identification. We added a fourteenth source reported by \cite{abeysekara2018} by their own reprocessing of Fermi-LAT data.
{Table\,\ref{tab:fermisources} lists name, central position, extension, identified or  likely associated  source/counterpart, reference and variability index from 3FGL and 4FGL catalogs, of the}  GeV discrete sources considered here. We verified that the 4FGL-DR2 contains the same sources as DR1, and used the error ellipses of DR2. 
Figure\,\ref{fig:observedregion} shows the observed area and the location and extension of the studied sources.

\begin{deluxetable*}{l r r l c l r}
\tablecaption{Discrete sources detected by Fermi in the observed Cygnus region\label{tab:fermisources}}
\tablewidth{0pt}
\tablehead{
 \colhead{Name} & \colhead{$RA_{\rm J2000}$}  & \colhead{$Dec_{\rm J2000}$} & \colhead{$\theta_1$, $\theta_2$, $PA^{\rm a}$}  & \colhead{Assoc. source$^{\rm b}$} & \colhead{Reference} & \colhead{Var.}\\
 & \colhead{(hms)} & \colhead{(dms)} & \colhead{(deg, deg, deg)} & & & index
}
\startdata
FGL\,J2018.1$+$4111 & 20:18:07.4 & $+$41:10:44 & 0.1, 0.1$^{\rm c}$ & --- & AB2018 & \\
3FGL\,J2018.6$+$4213 & 20:18:41.89 &  $+$42:13:49.4 & 0.212, 0.112, $+$57 & --- & AC2015 & 35.95\\ 
4FGL\,J2021.5$+$4026 & 20:21:32.4 & $+$40:26:40 & 0.01, 0.01, $-$31.7 &  PSR\,J2021$+$4026 & AB2020 & 201.76\\ 
  \,\,\,\,3FGL\,J2021.5$+$4026 & 20:21:32.5 & $+$40:26:51.68 & 0.008, 0.008, $+$17 & & AC2015 &  157.71\\
4FGL\,J2023.4$+$4127 & 20:23:29.1 & $+$41:27:03 &  0.03, 0.03, $+$40.0 &  SSTSL2 J202336.19$+$ & AB2020 &  11.03 \\ 
  \,\,\,\,3FGL J2023.5+4126 & 20:23:30.19 &  $+$41:26:01.28 &  0.098, 0.09, $+$85 & $+$412527.0 & AC2015 &  48.95\\
3FGL\,J2026.8$+$4003 & 20:26:51.79 & $+$40:03:09.28 & 0.228, 0.112, $+$82 & --- & AC2015 &  55.73 \\
3FGL\,J2028.5$+$4040c & 20:28:32.29 & $+$40:40:37.19 & 0.143, 0.093,  $-$29 & ---& AC2015 &48.79 \\
4FGL\,J2030.9$+$4416 & 20:30:54.1 & $+$44:16:01 & 0.02, 0.01, $-$86.0 &  PSR\,J2030$+$4415 & AB2020 &3.90\\ 
  \,\,\,\,3FGL\,J2030.8$+$4416 & 20:30:51.70 & $+$44:16:13.40 & 0.039, 0.038, $+$22 & & AC2015  & 36.43\\
4FGL\,J2032.2$+$4127 & 20:32:15.0 & $+$41:27:32 & 0.01, 0.01, $+$59.2 & PSR\,2032$+$4127 & AB2020 & 10.65\\ 
  \,\,\,\,3FGL\,J2032.2$+$4126 & 20:32:14.29 & $+$41:26:48.8 & 0.014, 0.014, $+$17 & & AC2015 &  38.32 \\
4FGL\,J2032.6$+$4053 & 20:32:36.6 & $+$40:53:38 & 0.17, 0.09, $-$63.4 & Cyg\,X-3 & AB2020 & 67.44\\ 
3FGL\,J2032.5$+$4032 & 20:32:30.0 & $+$40.32:05.6 & 0.165, 0.115, $+$7 & --- & AC2015 & 141.31 \\
3FGL\,J2036.8$+$4234c & 20:36:53.2 &  $+$42:34:05.08 & 0.077, 0.073, $+$84 & --- & AC2015 & 42.08 \\ 
3FGL\,J2037.4$+$4132c & 20:37:24.89 & $+$41:32:02.39 & 0.09, 0.07, $+$25 & --- &AC2015 & 40.87 \\ 
4FGL\,J2038.4$+$4212 & 20:38:30.0 & $+$42:12:31 & 0.05, 0.04, $-$14.3 & --- & AB2020 & 5.52\\ 
   \,\,\,\,3FGL\,J2038.4$+$4212 & 20:38:29.89 & $+$42:12:30.6 & 0.059, 0.054, $+$14 &--- & AC2015 & 45.67\\
3FGL\,J2039.4$+$4111 & 20:39:24.96  & $+$41:11:52.8 & 0.2, 0.1, $-$59 & --- & AC2015 & 45.39  \\ 
\enddata
\tablecomments{a: $\theta_1$, $\theta_2$ and $PA$  are the major and minor axis and  position  angle  of the Fermi error ellipse, at 68\% for 4FGL sources. b: Identified or likely associated source. c: The authors quote a point source and a PSF of 0.1~deg.  AB2018: \cite{abeysekara2018}; AC2015: \cite{acero2015}; AB2020: \cite{abdollahi2020}. 
}
\end{deluxetable*}

\subsection{Other gamma-ray sources}

Beside the discrete sources detected by the Fermi LAT telescope, there are other gamma-ray sources that lay in the observed region. This second group includes extended sources reported under the Fermi programs, and sources detected at TeV  energies, using other instruments, like the Cherenkov instruments High-Energy-Gamma-Ray Astronomy (HEGRA), High Altitude Water Cherenkov (HAWC), and the Very Energetic Radiation Imaging Telescope Array System (VERITAS).
Their names, position and extension are given in  Table\,\ref{tab:tevsources} (see also  Figure\,\ref{fig:observedregion}).   
 
\begin{deluxetable*}{l r r r l}
\tablecaption{Other sources in the observed Cygnus region\label{tab:tevsources}}
\tablewidth{0pt}
\tablehead{
\colhead{Name} & \colhead{$RA_{\rm J2000}$}  & \colhead{$Dec_{\rm J2000}$} & \colhead{Size} &  \colhead{Reference}\\
& \colhead{(hms)} & \colhead{(dms)} & \colhead{(arcmin)} & 
}
\startdata
VER\,J2019$+$407 & 20:20:04.8& $+$40:45:36 & 17.4$\times$11.4 & AL2013 \\  
4FGL\,J2021.0$+$4031e & 20:21:04.8& $+$40:31:12 & 37.8 & AB2020\\ 
4FGL\,J2028.6$+$4110e & 20:28:40.8 &$+$41:10:11.9 & 180 & AB2020\\
eHWC\,J2030$+$412 & 20:30:57.6& $+$41:13:48 & 10.8 &  ABE2020  \\
VER\,J2031$+$415 & 20:31:33.8 &$+$41:34:38.4 & 9.5$\times$4 & BA2014\\
TeV\,J2032$+$4130 & 20:32:07 &$+$41:30:30 & 11.2 & AH2002 \\
\enddata
\tablecomments{AL2013: \cite{Aliu2013}; AB2020: \cite{abdollahi2020}; ABE2020: \cite{abey2020PhRvL}; 
BA2014: \cite{bartoli2014}, AH2002: \cite{Aharonian2002}.}
\end{deluxetable*}

\section{Data calibration and imaging} \label{sec:datacal}

The radio observations used for this investigation  were carried out with the GMRT at the bands centred at 325
and 610~MHz, during four campaigns between 2013 and 2017, with a total of 172 hours of observing time.  
The field of views (FoV) of the GMRT are 81$\pm$4$'$ and 43$\pm$3$'$ (GMRT Observer’s
Manual\footnote{www.ncra.tifr.res.in/ncra/gmrt/gmrt-users /observing-help})
at 325~MHz and 610~MHz bands respectively. 
According to the FoVs sizes and the region to cover (see Figure\,\ref{fig:observedregion}),   
we implemented five pointings at 325~MHz and 47 pointings at 610~MHz \citep[45 plus repeating two additional with originally bad data; again, see Figure\,2 of][]{benaglia2020b}, and taking also into account the requirement to obtain a uniform noise with the minimum amount  of pointings.
The flux calibrators 3C286 and/or 3C48 were observed at the beginning and at the end of the observing session. A secondary calibrator 2052$+$365 was observed for 5 minutes after every 30 mins scan on the target. To minimise the effect of bandwidth smearing, the signal with 33.33 MHz bandwidth was recorded in spectral line mode with 256 channels.

The data was processed with the Source Peeling and Atmospheric Modeling algorithms \citep{intema2014}, that handle both the calibration and  the imaging steps. Full description of acquisition, reduction and imaging of the data is presented with all details in \cite{benaglia2020b}. At imaging, we adopted a robust weighting equal to $-$1. The synthesised beams of the final mosaics were $10'' \times 10''$ for the 325 MHz band, and $6'' \times 6''$ at the 610 MHz band. The attained rms values resulted, on average, of up to 0.5~mJy~beam$^{-1}$ and 0.2~mJy~beam$^{-1}$, respectively. The final individual images were mosaicked to get a single image of full region.

Additionally, we also imaged the  325~MHz FoVs, with the Astronomical Imagining Processing System (AIPS) following standard procedures. The data were inspected and flagged manually. The flux density scale for the flux calibrators 3C48 and 3C286 was set using Perley-Butler 2013 scale \citep{PerleyButler2013}.  After the bandpass calibration, the data were averaged keeping in mind to minimise the effect of bandwidth smearing. Multi-facet imaging options were used to correct for w-term due to large field of view. A few rounds of phase-only self-calibration were carried out to correct for phase variations, which has improved the image quality substantially. The primary beam correction was carried out on the final image.  The $T_{\rm sys}$ correction due to excess background emission at low radio frequencies in the galactic plane was also applied on the final images.

\section{Results} \label{sec:results}

\subsection{Radio emission  in discrete Fermi sources}

We inspected the images at 325 and 610~MHz to search for radio sources above 3$\sigma$ (rms), that laid in the error ellipses of the GeV sources listed in Table\,\ref{tab:fermisources}.
In the process, signals that represented peaks of extended and/or diffuse emission were discarded. For nine discrete Fermi objects, 35 radio sources were found, five of them at only one band.

To derive the radio flux densities, we convolved the 610~MHz image to the synthesised beam of $10''$. We fit Gaussian functions, and verified the value of the  integrated flux obtained in that way, by measuring the flux density above the 3$\sigma$ contour, being $\sigma$ the local rms. The measurements are given in Table\,\ref{tab:fluxs-spixs}. 

For the 30 radio sources detected at both bands, we derived the spectral index $\alpha$, using the convention $S \propto \nu^\alpha$. For the rest, a spectral index upper limit is quoted (see Table\,\ref{tab:fluxs-spixs}), except for one with observations only at one band  (S35). One must take into account here that the observations at the two radio bands were not simultaneous \citep[see Table\,2 of ][]{benaglia2020b}, and that by using the mosaic technique adjacent FoVs limiting areas were averaged.

In what follows, we describe the findings related to each gamma-ray source of Table\,\ref{tab:fermisources}. The individual images of the 
radio sources are presented in the Appendix.

\begin{figure}[ht!]
\centering
\includegraphics[width=8.22cm]{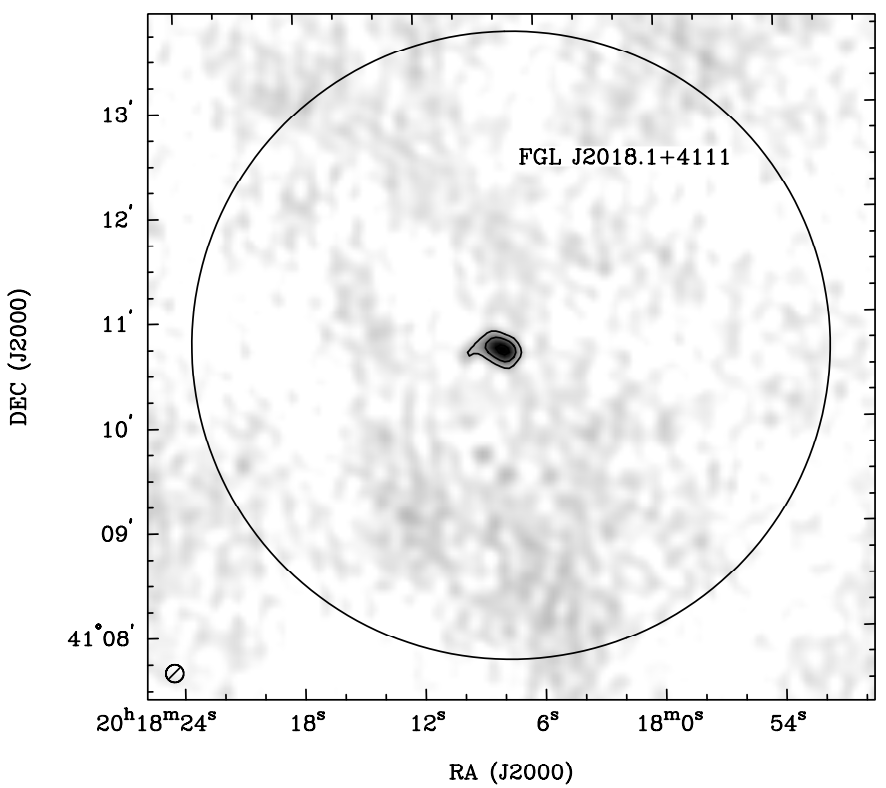}
\includegraphics[width=8cm]{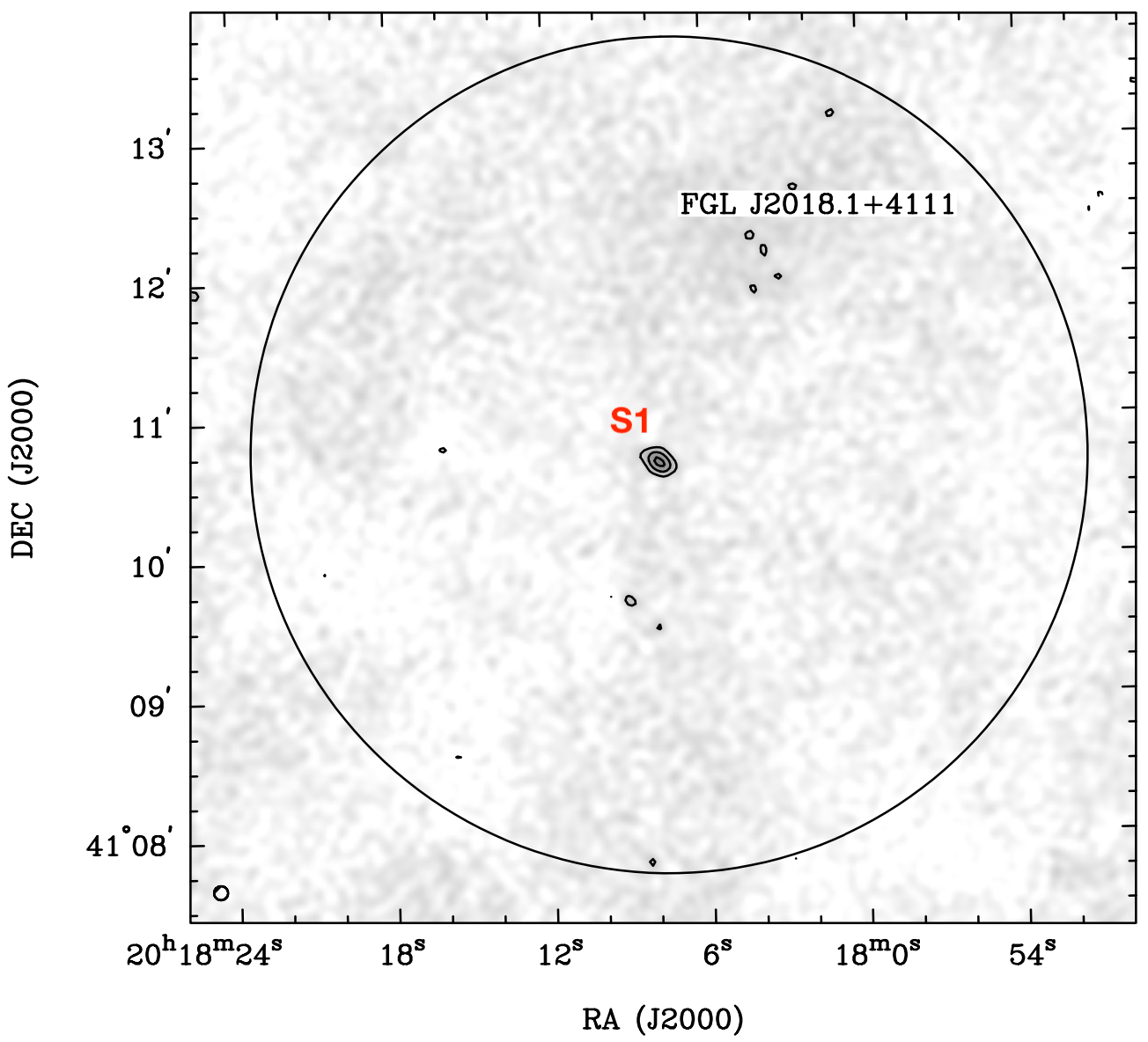}
\caption{FGL\,J2018.1$+$4111. Left: 325~MHz emission; contour levels at $-$3, 3, 7 and 12 in units of $\sigma$ (0.25~mJy beam$^{-1}$). Right: 610~MHz emission; contour levels at $-$0.3, 0.3, 6 and 8.5 in units of $\sigma$ (0.1~mJy beam$^{-1}$). The central source is identified as S1. The  synthesized beam is shown in the bottom right corner.}
\label{fig:fgl2018}
\end{figure}

\begin{figure}[h!]
\centering
\includegraphics[width=8cm]{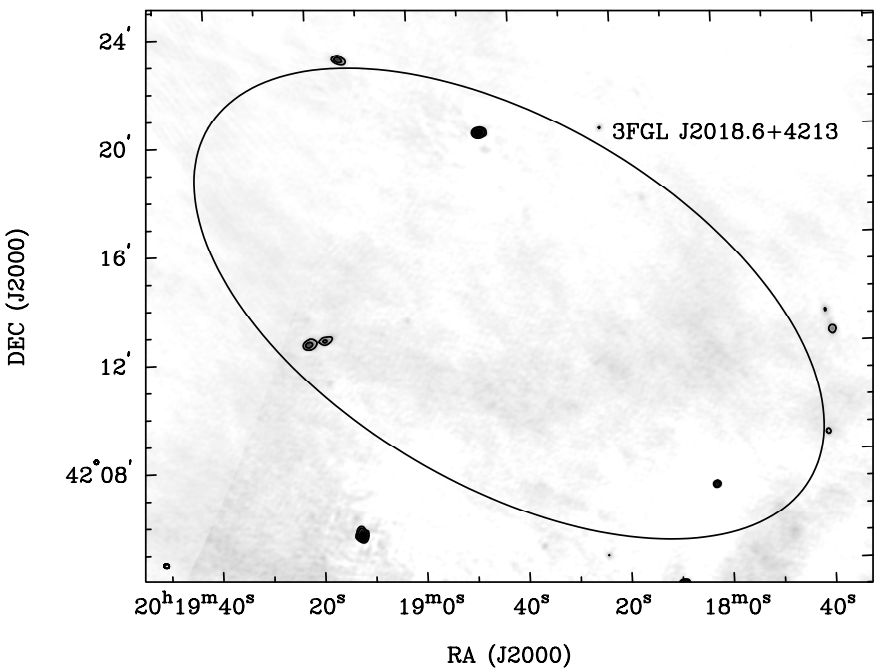}
\includegraphics[width=8.22cm]{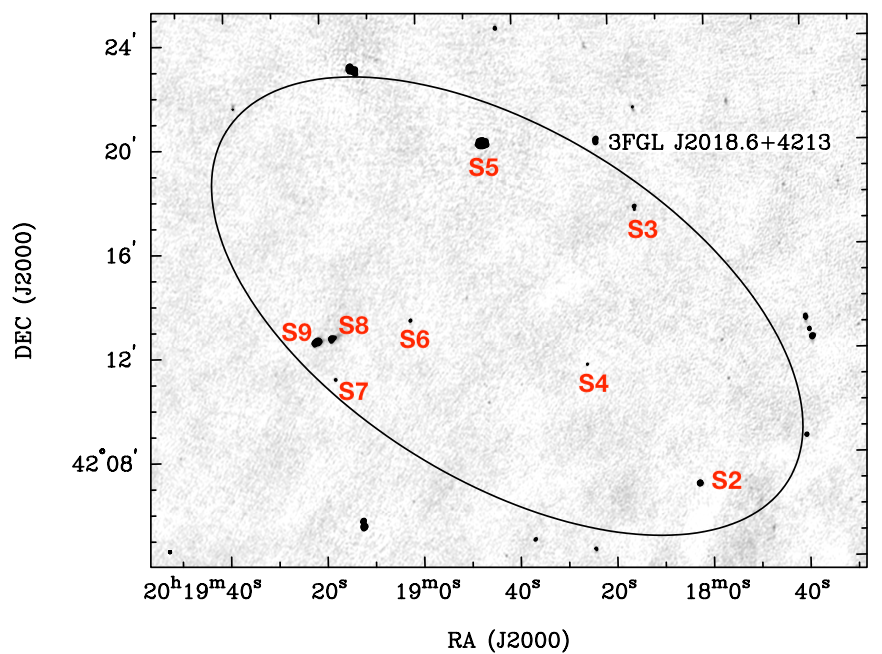}
\caption{3FGL\,J2018.6$+$4213. Left: 325~MHz emission; contour levels at $-$3, 3 and 7 in units of $\sigma$ (1.5~mJy beam$^{-1}$). Right: 610~MHz emission; contour levels at $-$3, 3, 5 and 10 in units of $\sigma$ (0.35~mJy beam$^{-1}$). 
The radio sources S2 to S9 are identified.}
\label{fig:3j20186}
\end{figure}

\begin{figure}[h!]
\centering
\includegraphics[width=8cm]{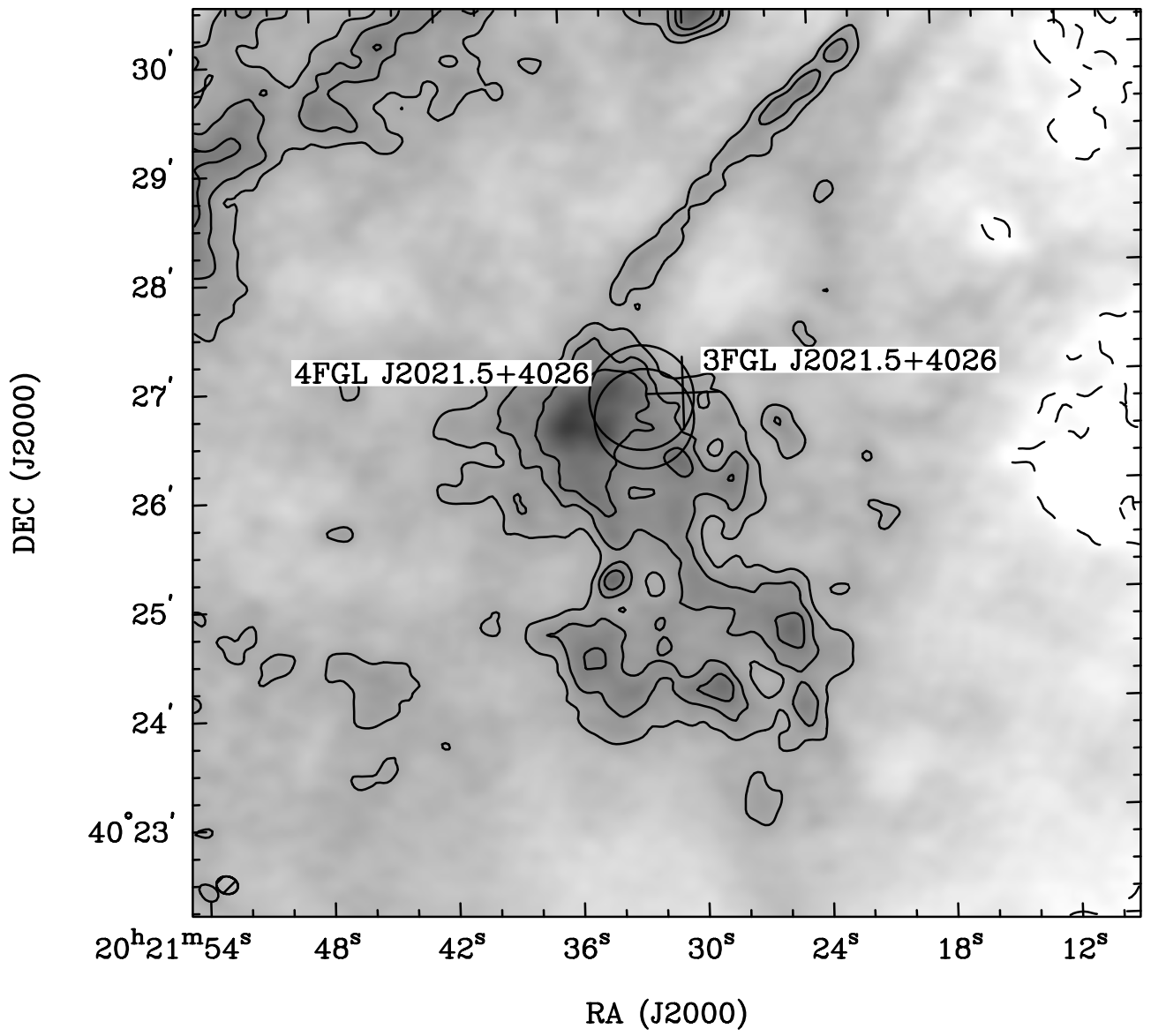} 
\includegraphics[width=8cm]{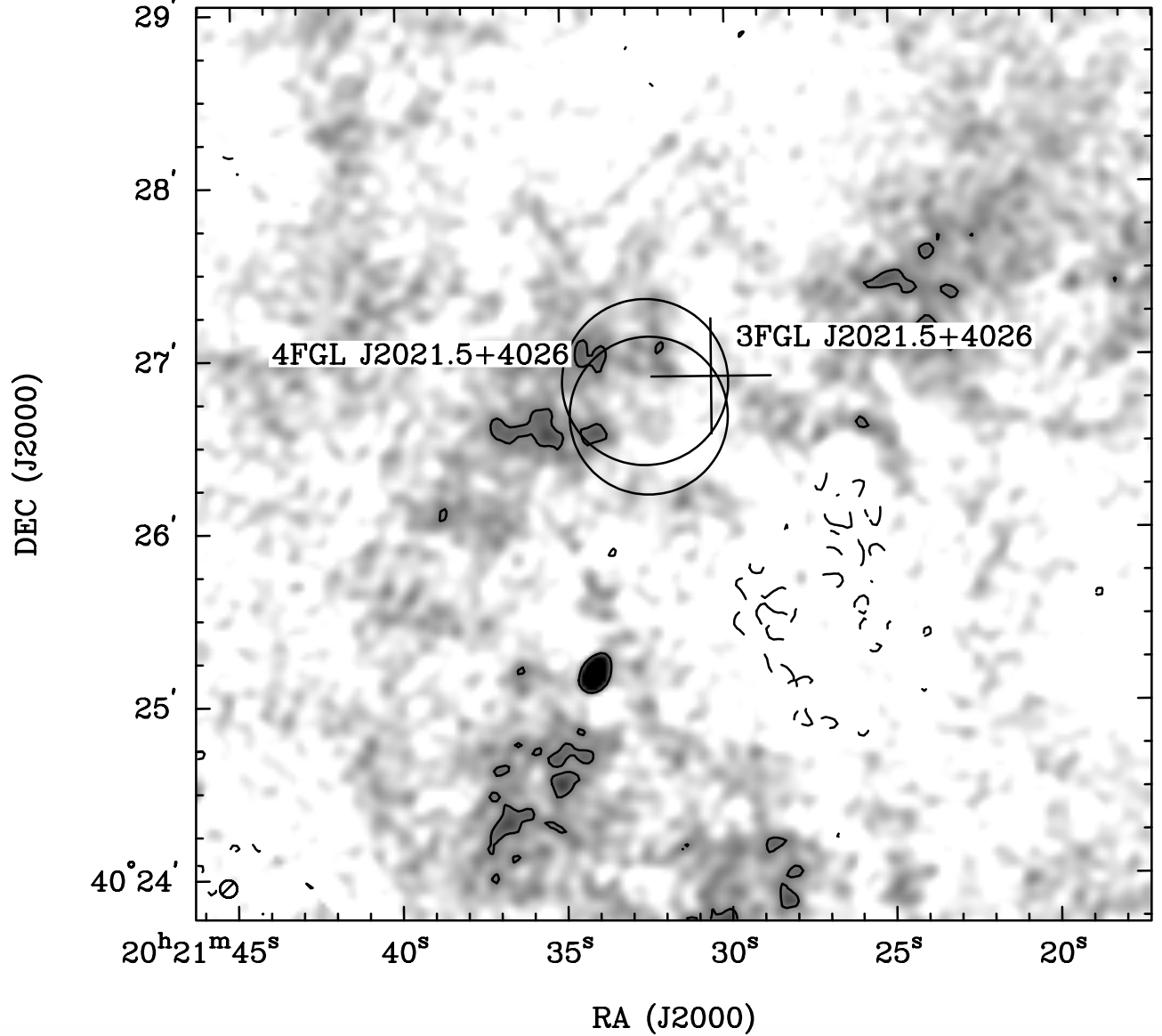} 
\caption{4FGL~J2021.5$+$4026, represented by the  southern circle, at 68\%. Left: gray scale emission at 325~MHz with contour levels -2.5, 2.5, 3.75 and 5~mJy beam$^{-1}$. Right: emission at 610~MHz with contour levels at $-$1, 1, 
3 and 5~mJy beam$^{-1}$. The northern circle marks 3FGL~J2021.5$+$4026, and the cross marks the position of the pulsar PSR~J2021$+$4026.}
\label{fig:4j20215}
\end{figure}

\begin{figure}[ht!]
\centering
\includegraphics[width=8cm]{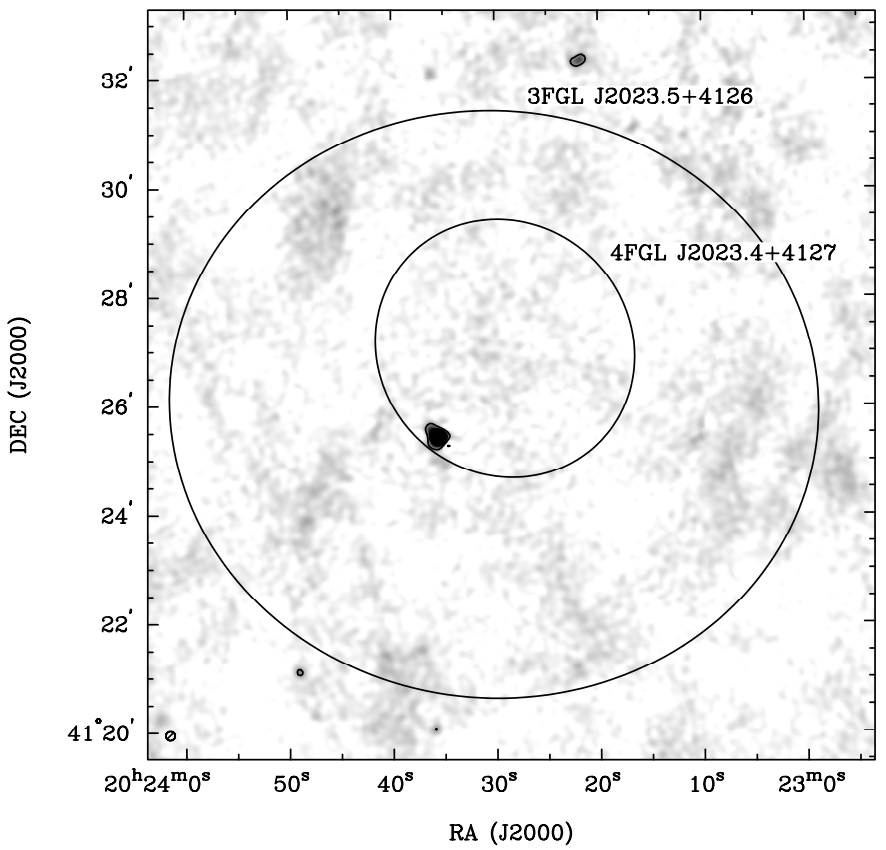}
\includegraphics[width=8cm]{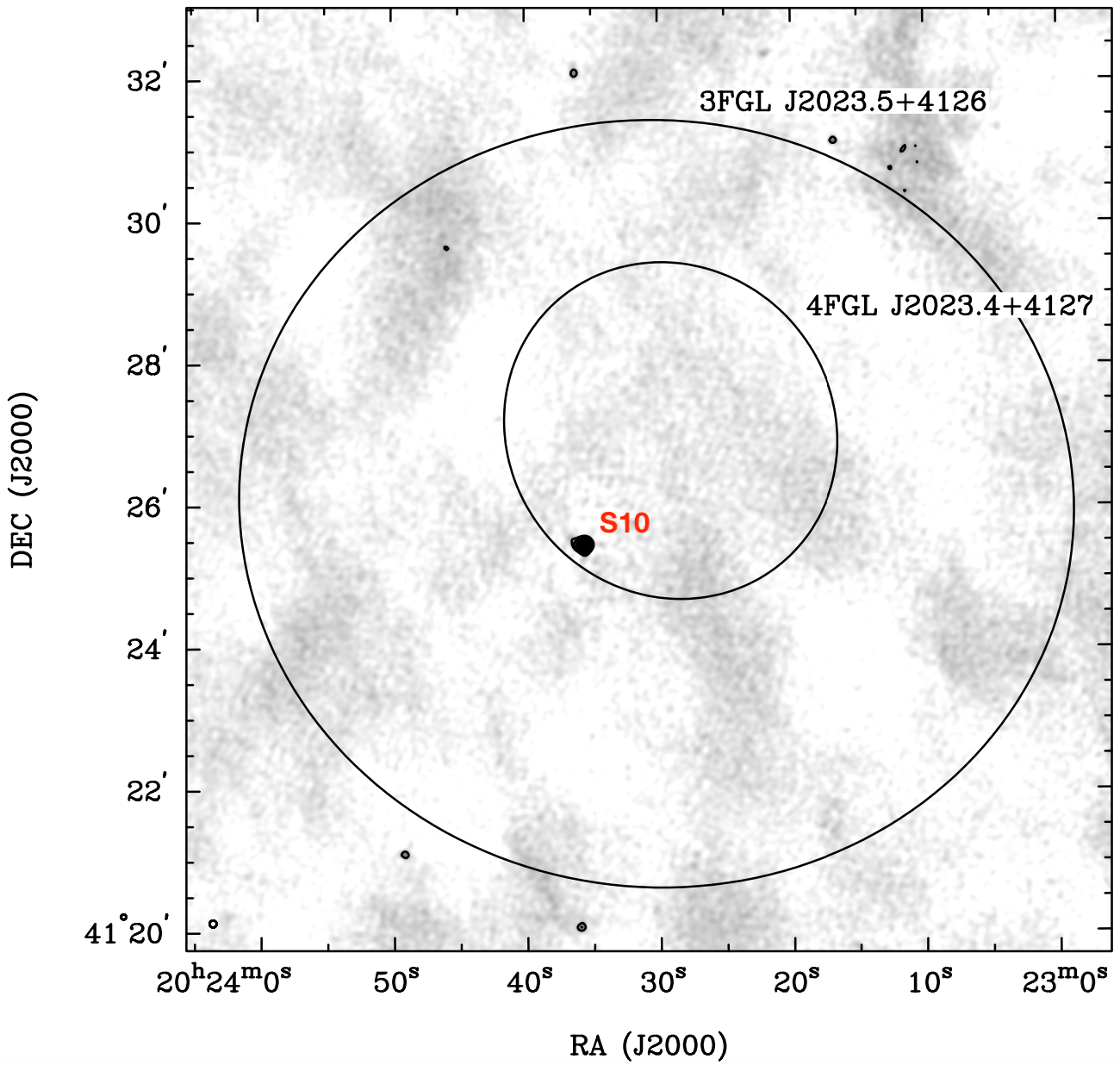}
\caption{4FGL~J2023.4$+$4127 at 95\%. Left: 325~MHz  emission; contour levels at $-$3, 3, 10 and 20 in units of $\sigma$ (0.5~mJy beam$^{-1}$). Right: 610~MHz emission; contour levels $-$3, 3, 6 and 30 in units of $\sigma$ (0.2~mJy beam$^{-1}$). The largest circle represents the position of 3FGL~J2023.5$+$4126. The radio source S10 is identified.}
\label{fig:4j20234}
\end{figure}

{\bf FGL\,J2018.1$+$4111} was discovered by \cite{abeysekara2018} and reported as a point source of unknown nature, and a PSF at 1~TeV of 0.1~deg.
Only one { compact} radio source at the exact central position  of this Fermi source was detected in the radio images at both bands, and named S1; see  Figure~\ref{fig:fgl2018}. 
The spectral index between 325 and 610~MHz resulted in $\alpha = -1.6\pm0.2$. Such a steep radio spectrum is a strong indicator that this may be a pulsar.

In the  error  ellipse of {\bf 3FGL\,J2018.6$+$4213} 
we detected eight radio sources at  610~MHz above 3$\sigma$ emission (sources S2 to S9), and six of them also at 325~MHz (see Figure~\ref{fig:3j20186} and Table~\ref{tab:fluxs-spixs}), with flux  densities from  $\sim$1 to almost 500~mJy.
One of them is double (S3). Most spectral indices of the six detected at both bands are less or equal to $-1$, {but  one source (S6) exhibited a very steep index of $-2.1\pm0.5$. One possible reason for such a steep radio spectral index is the variability, with the flux variations such that to produce steep spectra. Another possibility is that this is a pulsar candidate. It will be useful to further confirm if the source is variable, if so, this may be a micro-quasar; some of them are VHE sources.}

We found no discrete radio sources at the locations of {\bf 4FGL\,J2021.5$+$4026}, 3FGL\,J2021.5$+$4026 or the pulsar PSR\,J2021$+$4026, but  diffuse and rather strong extended emission (see Figure~\ref{fig:4j20215}). Probably due to  that reason, the 610~MHz FoV (out  of the 47 pointings) in which the Fermi source is sitting, resulted with higher noise than average. 

The only radio source  at the  position of {\bf 4FGL\,J2023.4$+$4127} and 3FGL\,J2023.5$+$4126, S10, detected  at both radio bands, could be characterized with a spectral index of $+0.2\pm0.1$. 
We note that S10 is inside the 95\% error ellipse for this Fermi source; see Figure~\ref{fig:4j20234}.

Our images covered the northern half (at  325~MHz)  and the  western  half (at 610~MHz) of {\bf 3FGL\,J2026.8$+$4003}. No radio sources were found, neither at 325~MHz above 0.6~mJy~beam$^{-1}$, nor at 610~MHz FoVs above 1.0~mJy~beam$^{-1}$ of the observed area due to higher noise in this region.

\begin{figure}[h!]
\centering
\includegraphics[width=8cm]{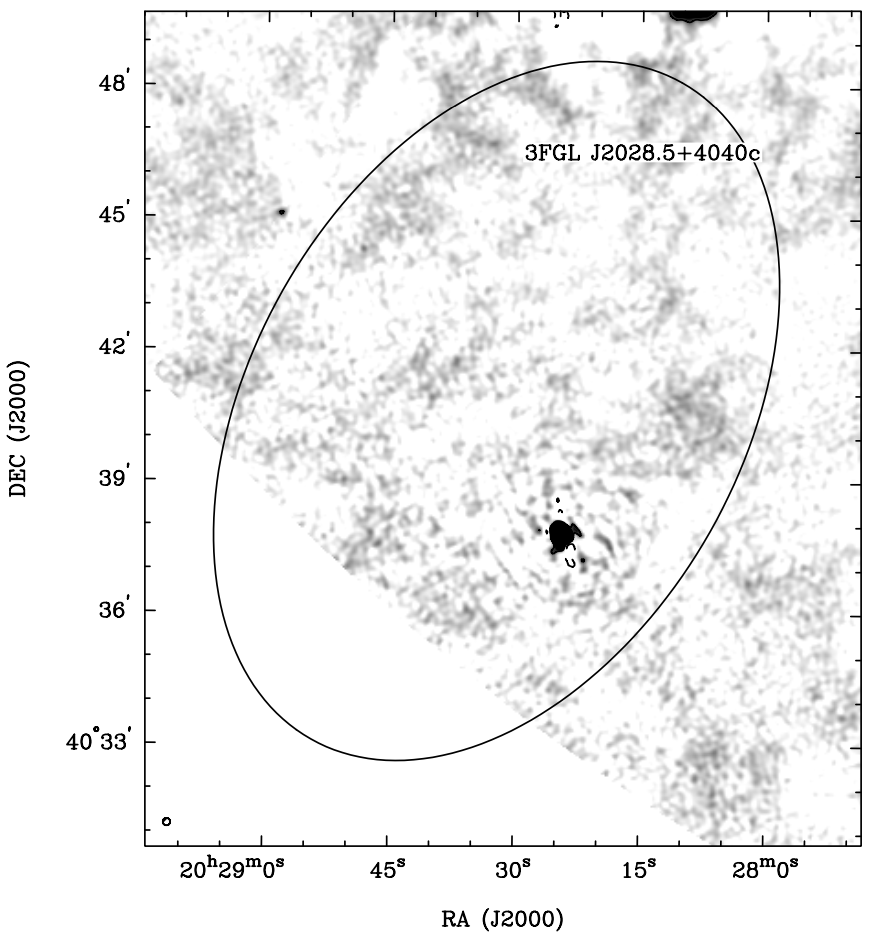}
\includegraphics[width=8cm]{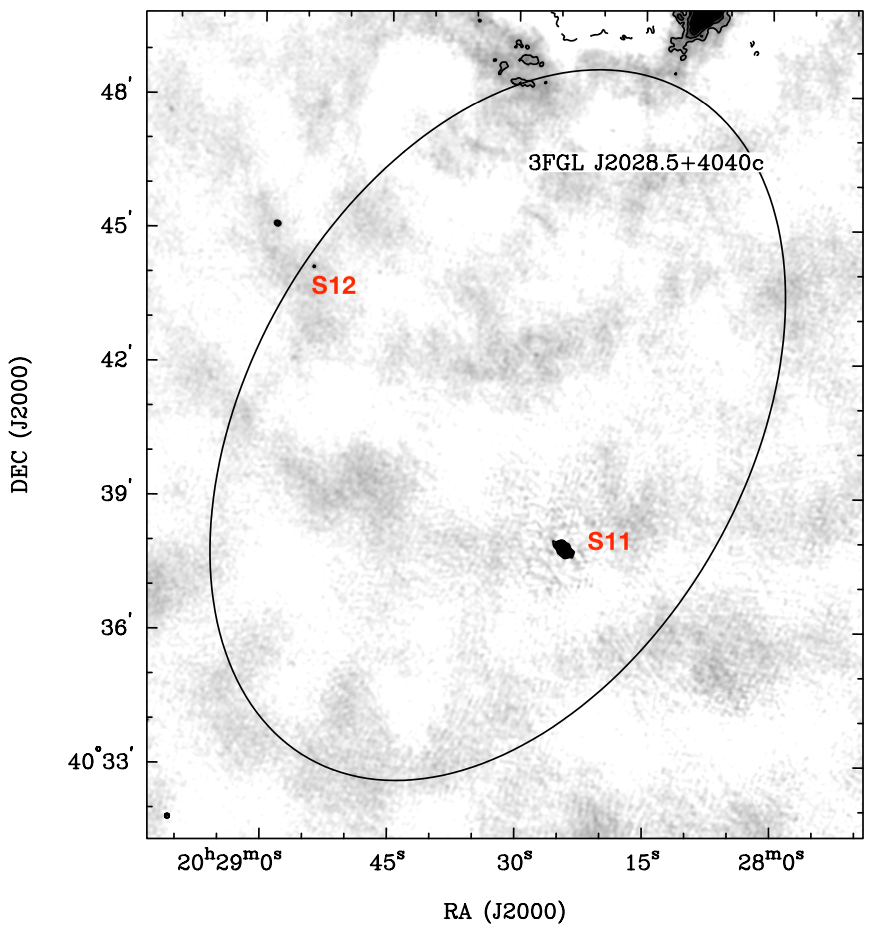}
\caption{3FGL\,J2028.5$+$4040c. Left: 325~MHz emission; contour levels at $-$3, 3, 5 and 10 in units of $\sigma$ (0.9~mJy beam$^{-1}$). Right: 610~MHz emission; contour levels at $-$3, 3, 5 and 7 in units of $\sigma$ (0.3~mJy beam$^{-1}$). The radio sources S11 and S12 are identified.}
\label{fig:3j20285}
\end{figure}

There are two radio sources in the area of {\bf 3FGL\,J2028.5$+$4040c}. 
The  brighter ($S/N>350$), { S11, double at 610~MHz}, presents a  spectral index of $-1.2\pm0.1$. The other was detected solely  at 610~MHz (see Figure~\ref{fig:3j20285} and Table~\ref{tab:fluxs-spixs}).

\begin{figure}[ht!]
\centering
\includegraphics[width=8cm]{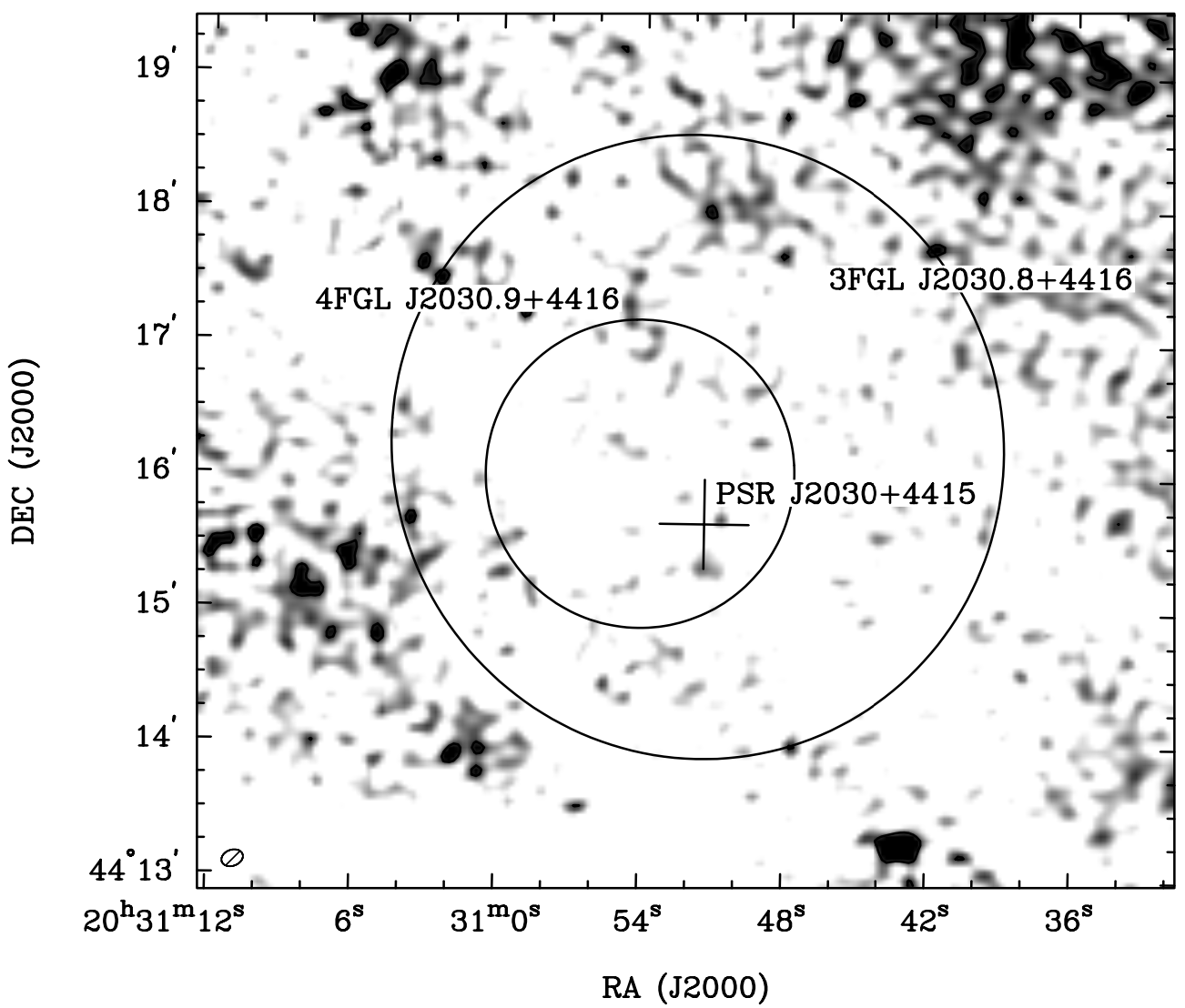}
\includegraphics[width=8cm]{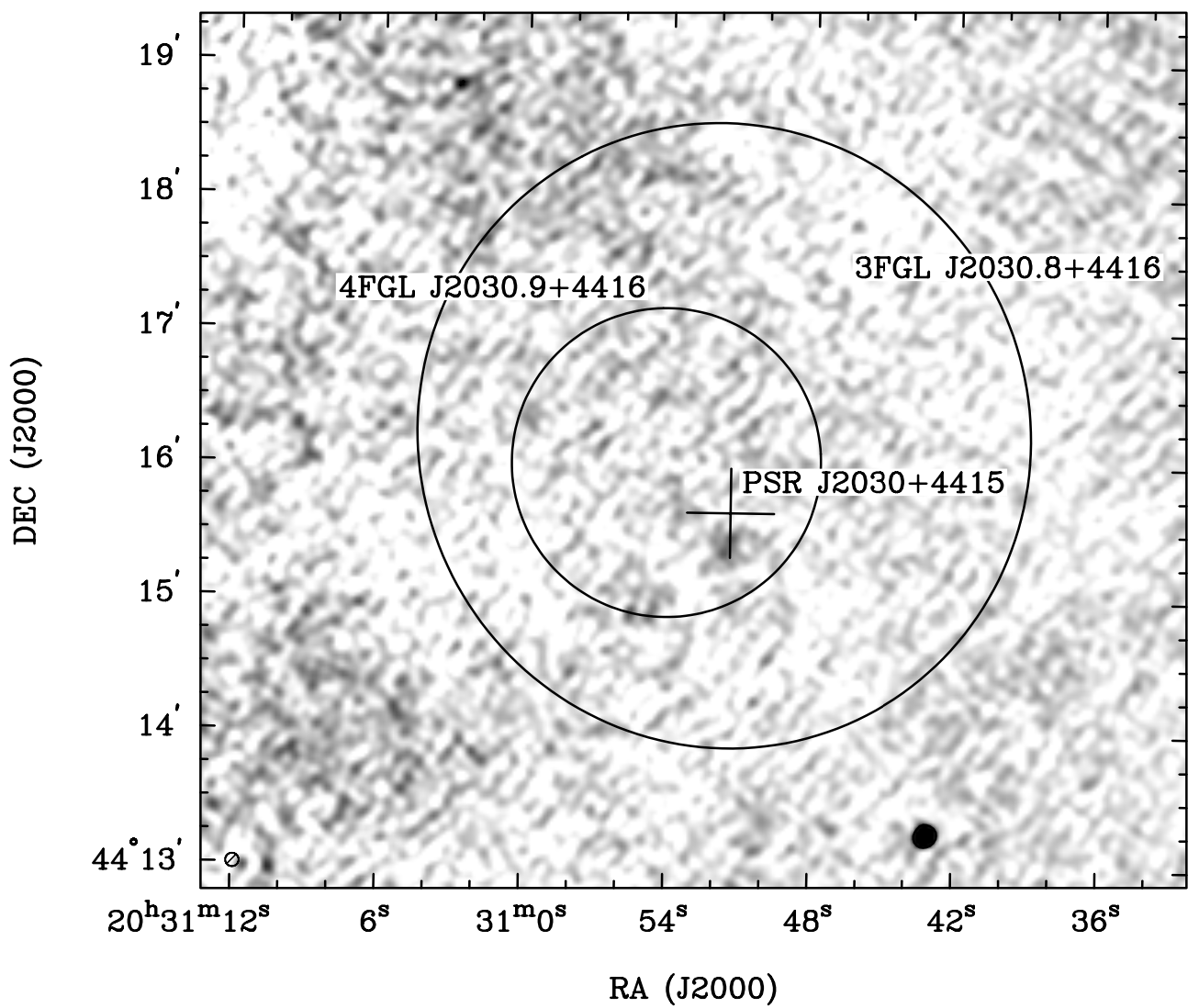}
\caption{4FGL\,J2030.9$+$4416. Left: 325~MHz emission; contour level at 2, in units of $\sigma$ (0.6~mJy beam$^{-1}$). Right: 610~MHz emission; contour levels at $-$2, and 2 in units of $\sigma$ (0.5~mJy beam$^{-1}$). The position of 3FGL~J2030.8$+$4416 is also represented. The cross marks the position of the pulsar PSR~J2030$+$4415.}
\label{fig:4j20309}
\end{figure}

\begin{figure}
\centering
\includegraphics[width=8cm]{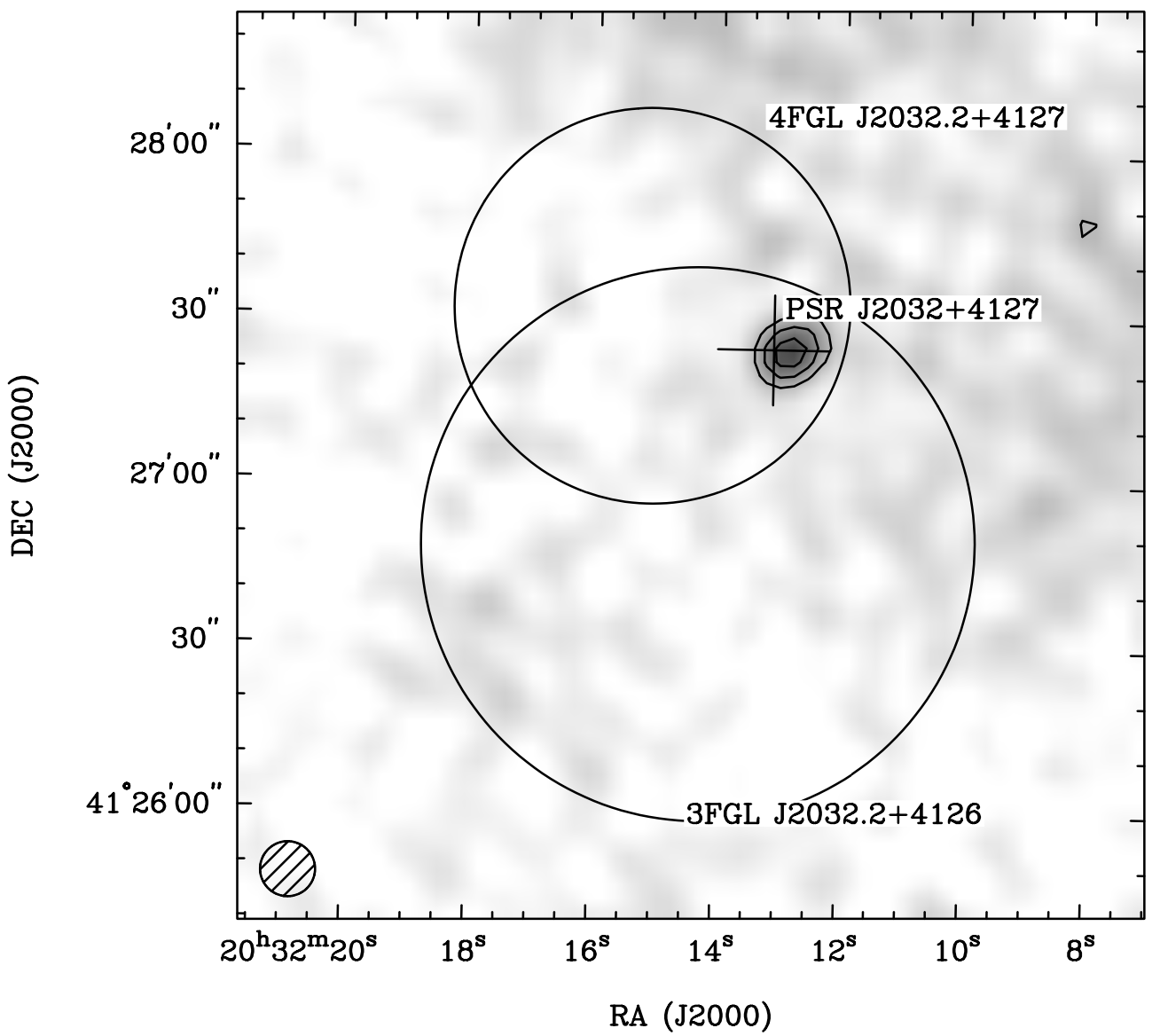}
\includegraphics[width=7.6cm]{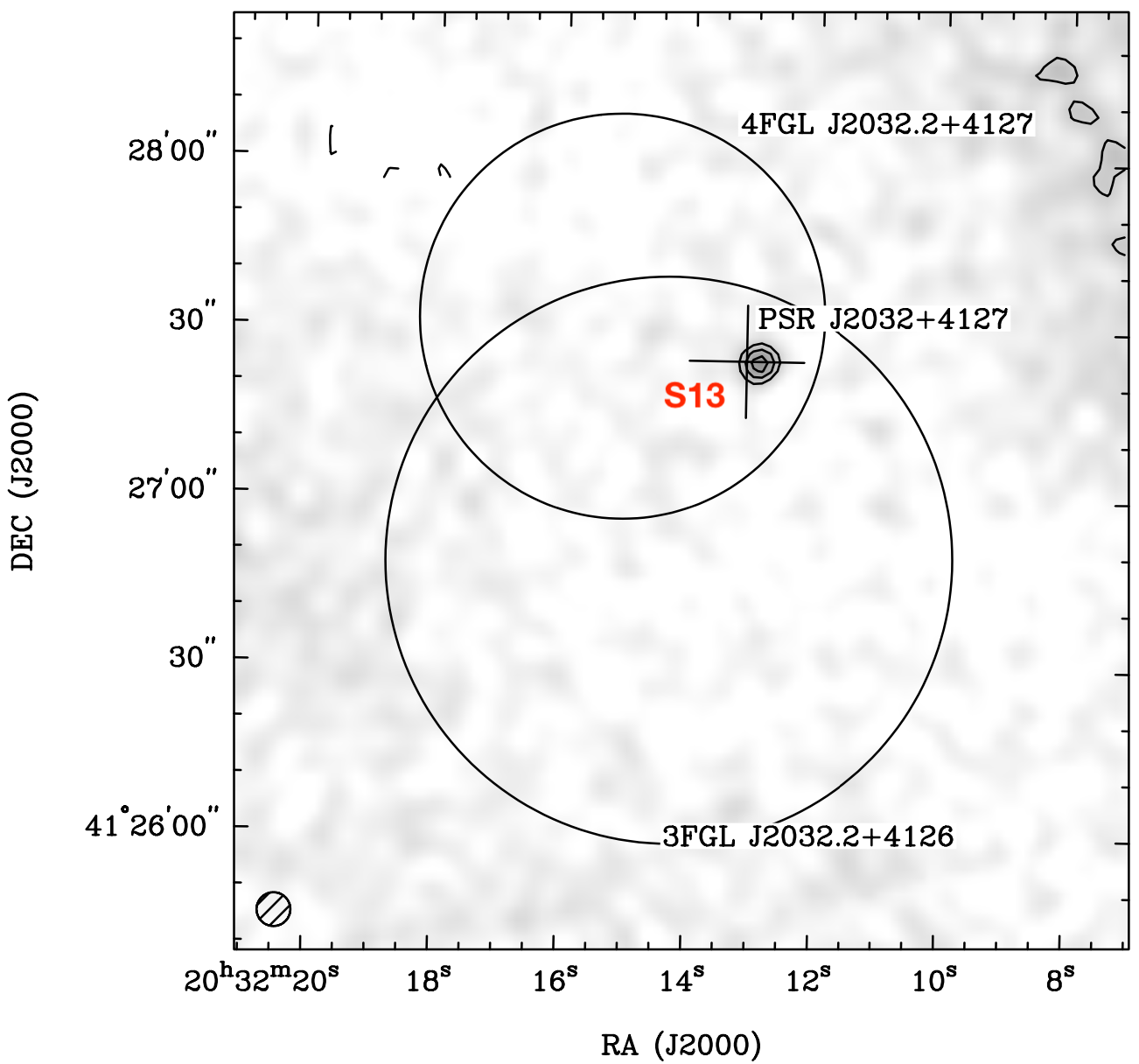}
\caption{4FGL~J2032.2$+$4127. Left: 325~MHz emission; contour levels at $-3$, 3, 5 and 7 in units of $\sigma$ (0.3~mJy beam$^{-1}$). Right: 610~MHz emission; contour levels $-$3, 3, 5, and 7~mJy in units of $\sigma$ (0.1~mJy beam$^{-1}$). The position of 3FGL~J2032.2$+$4126 is also represented. The cross marks the position of the pulsar PSR~J2032$+$4127. The radio source S13 is identified.}
\label{fig:4j20322}
\end{figure}

\begin{figure}
\centering
\includegraphics[width=8.5cm]{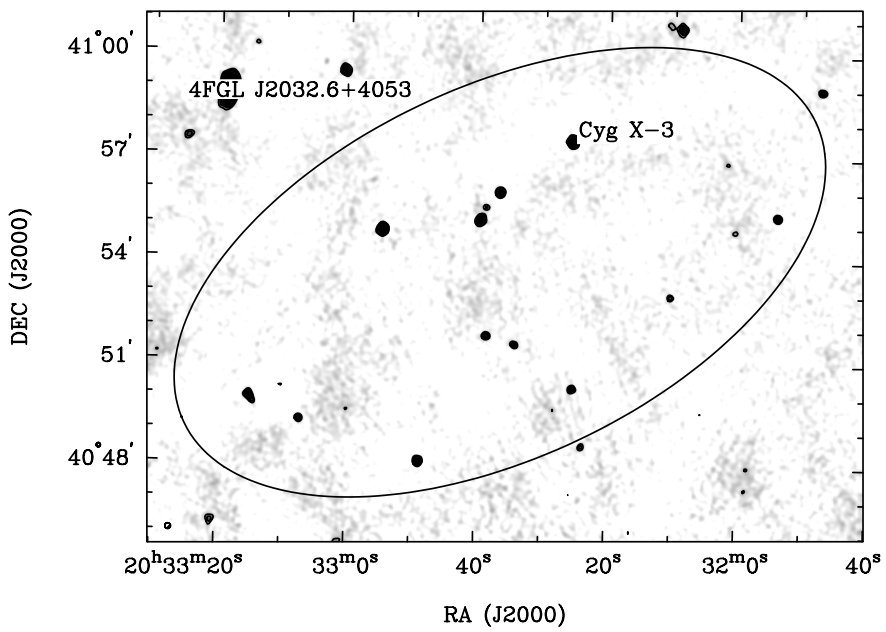}
\includegraphics[width=8.5cm]{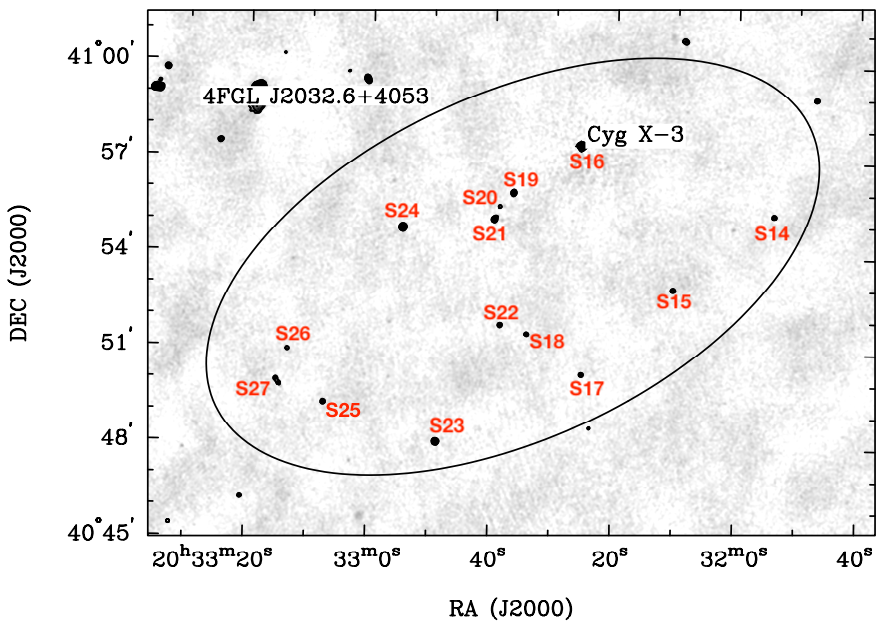}
\caption{4FGL~J2032.6$+$4053. Left: 325~MHz emission; contour levels $-3$, 3, 5, and 7 in units of $\sigma$ (0.4~mJy~beam$^{-1}$). Right: 610-MHz emission; contour levels $-3$, 3, 6, 10, and 15 in units of $\sigma$ (0.2~mJy~beam$^{-1}$). The radio sources S14 to S27 are identified, together with the object Cyg\,X-3.}
\label{fig:4j20326}
\end{figure}

\begin{figure}
\centering
\includegraphics[width=8cm]{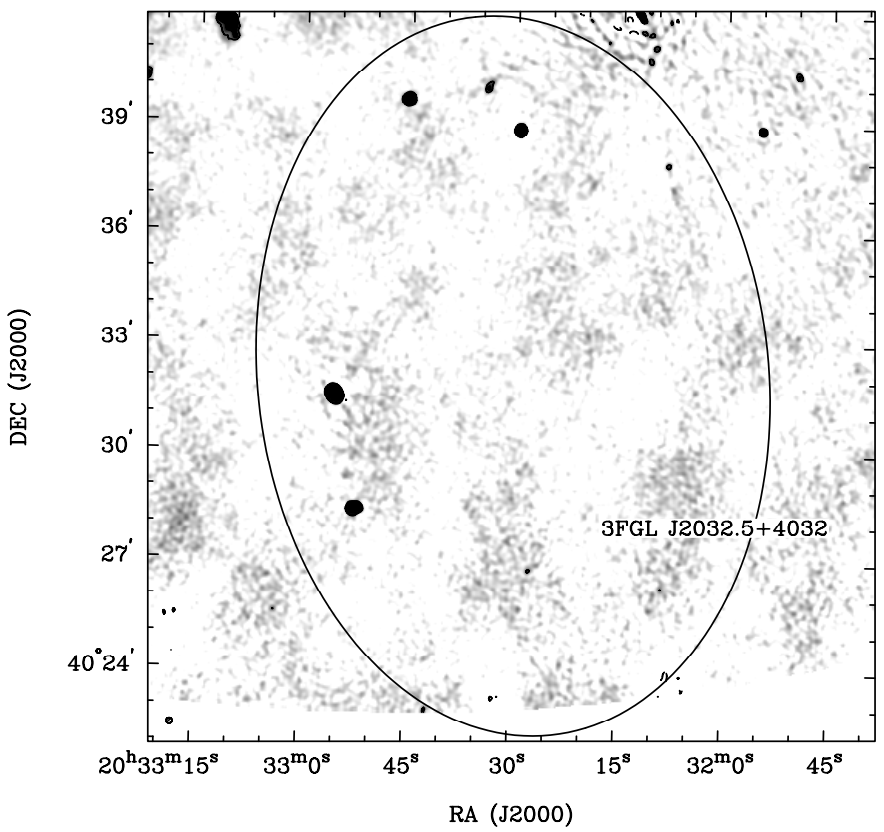}
\includegraphics[width=7.7cm]{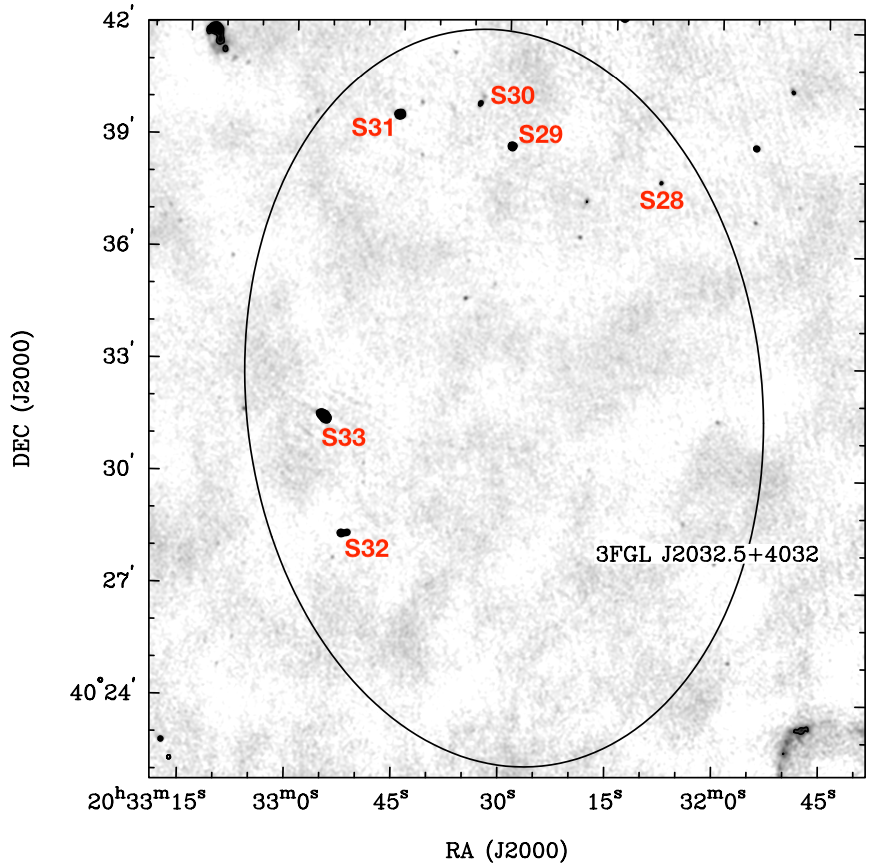}
\caption{3FGL\,J2032.5$+$4032. Left: 325~MHz emission; contour levels $-3$, 3 and 5 in units of $\sigma$ (0.7~mJy beam$^{-1}$). Right: 610-MHz emission; contour levels $-3$, 3 and 5 in units of $\sigma$ (0.45~mJy beam$^{-1}$). The radio sources S28 to S33 are identified.}
\label{fig:3j20325}
\end{figure}
 
\begin{figure}
\centering
\includegraphics[width=7.6cm]{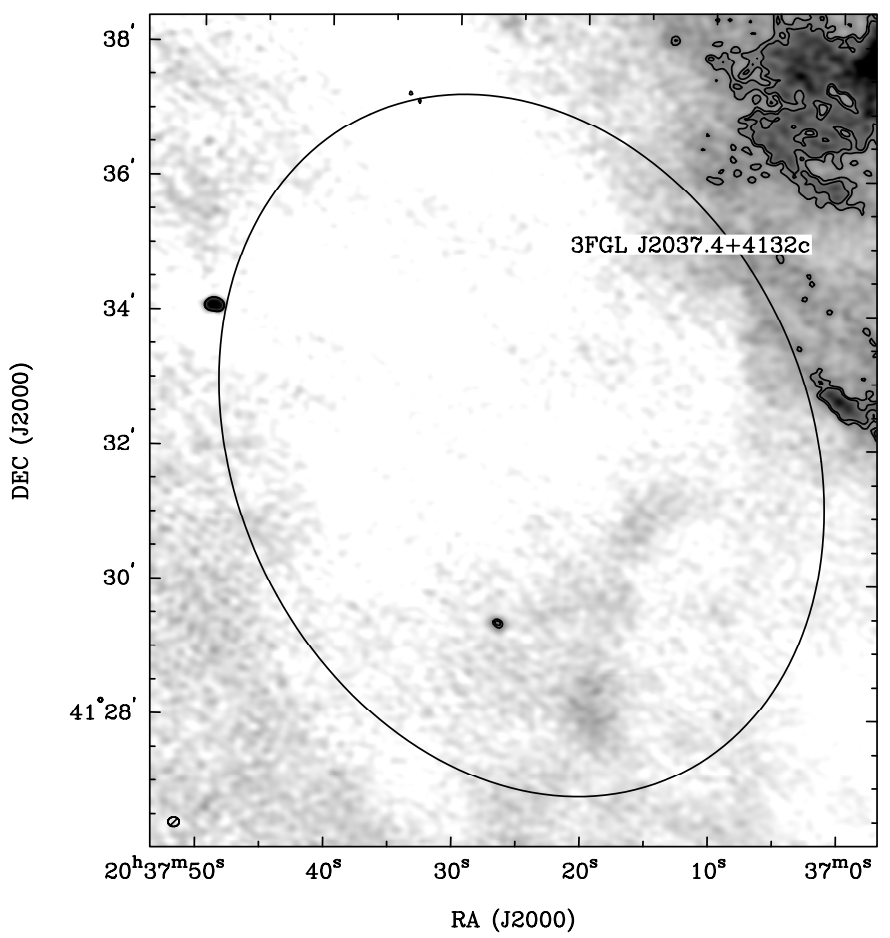}
\includegraphics[width=8cm]{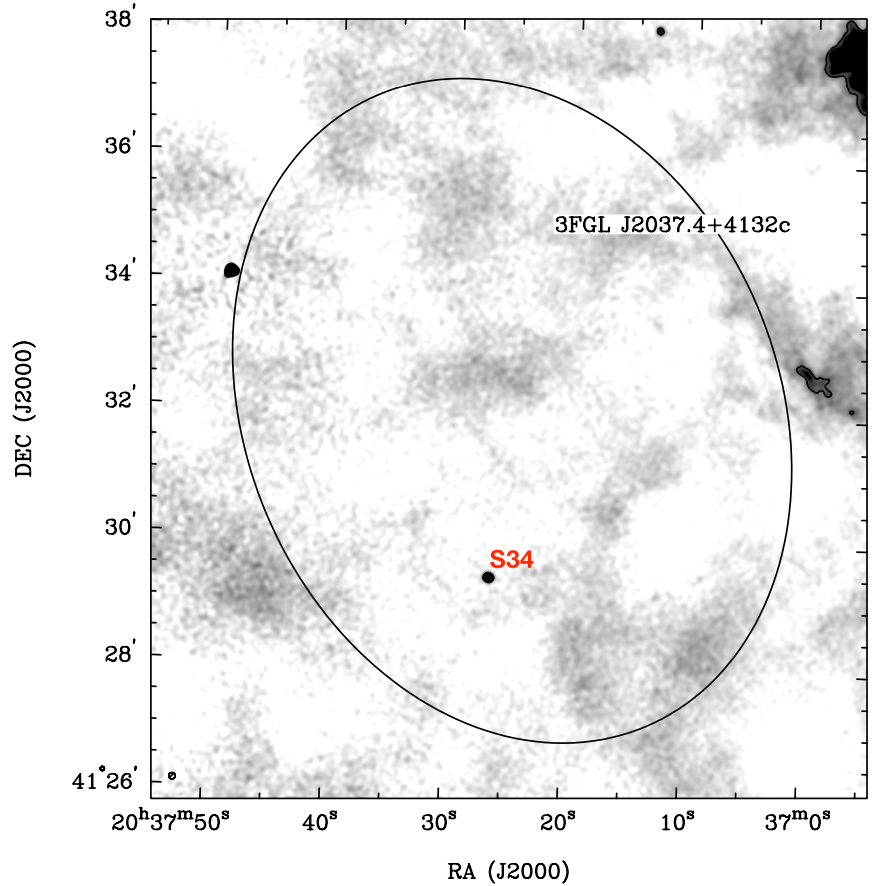}
\caption{3FGL\,J2037.4$+$4132c. Left: 325~MHz emission; contour levels of 5 and  10 in units  of $\sigma$ (1~mJy beam$^{-1}$). Right: 610~MHz emission; contour levels $-3$, 3 and 5 in units of $\sigma$ (0.4~mJy beam$^{-1}$). The radio source S34 is identified.}
\label{fig:3j20374}
\end{figure}
 
\begin{figure}
\centering
\includegraphics[width=8cm]{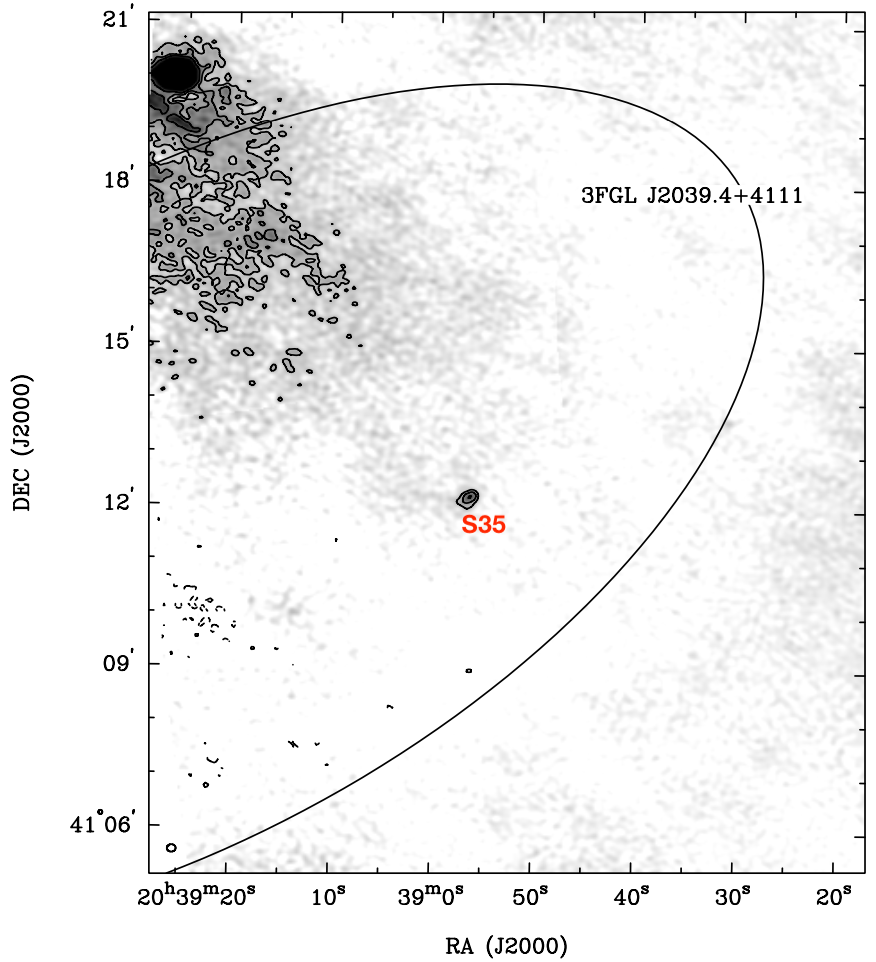}
\caption{3FGL\,J2039.4$+$4111. 325~MHz emission; contour levels $-3$, 3,  5 and 8 in units of $\sigma$ (1.6~mJy beam$^{-1}$). The radio source S35 is identified.}
\label{fig:3j20394}
\end{figure}

The area of {\bf 4FGL\,2030.9$+$4416} presents weak, diffuse radio emission, probably {resolved out} due  to the weighting scheme adopted in order to  outline point sources (see Figure~\ref{fig:4j20309}). There is an indication that part of that emission could be related to the object PSR\,J2030+4415 and its  wind nebulae \citep{devries2020}.
The pulsar wind nebulae is a potential candidate for source of VHE emission \citep{Bednarek2003}. Deeper multi-frequency radio observations are required to confirm this as pulsar wind nebulae which might be powering this VHE source.

In the case of {\bf 4FGL\,J2032.2$+$4127}, 
{there is a pulsar, namely PSR J2032+4127, which was detected by Fermi \citep{abdo} in gamma rays and
in radio with the Green Bank Telescope \citep{Camilo2009}. Our results showed} a radio source coincident with the position of this pulsar,
detected at both bands (see Figure~\ref{fig:4j20322} and Table~\ref{tab:fluxs-spixs}), and the spectral index derived was $-2.0\pm0.2$. { The steep radio spectral index of this known pulsar further strengthens the possibility that other very steep spectrum source may also be pulsars.}

Fourteen radio sources, S14  to S27, were detected inside the ellipse of {\bf 4FGL\,J2032.6$+$4053}, including one at the position of the micro-quasar Cyg\,X-3 (S16). Source S26 was only detected at 610~MHz (see Figure~\ref{fig:4j20326} and Table~\ref{tab:fluxs-spixs}), and S27 is double. Spectral indices range from $-1.6$ to  $+1.4$. 

The source {\bf 3FGL\,J2032.5$+$4032} overlaps six radio sources, detected at both bands with spectral indices between $-1.3$ and $+0.5$ (see Figure~\ref{fig:3j20325} and Table~\ref{tab:fluxs-spixs}). { Two of the radio sources, S32 and S33, are double.}

The ellipse of {\bf 3FGL\,J2036.8$+$4234c} was observed at 610~MHz, and  contained no radio sources above a 3$\sigma$ value of 0.4~mJy~beam$^{-1}$.

The sky area covered by {\bf 3FGL\,J2037.4$+$4132c} contains one radio source, S34, barely detected at both bands due to diffuse emission present in the field (see Figure~\ref{fig:3j20374} and Table~\ref{tab:fluxs-spixs}).

{\bf 4FGL\,J2038.4$+$4212:} only observed at 610~MHz,  no radio sources above a 3$\sigma$ value of 0.6~mJy~beam$^{-1}$.

The  observations at the 325~MHz band partly cover the ellipse of {\bf 3FGL\,J2039.4$+$4111}, where we detected a radio source tagged  as S35, and  some extended  emission towards the north-east (see Figure~\ref{fig:3j20394}).

\begin{deluxetable*}{{@{}l l r r r r l}}
\tablecaption{Flux density and spectral index of radio sources detected in the Fermi discrete sources\label{tab:fluxs-spixs}}
\tablewidth{0pt}
\tablehead{
\colhead{Gamma-ray} & \multicolumn{2}{c}{~~~~~~~~~~~Radio source} & \colhead{$S_{\rm 325MHz}$}  & \colhead{$S_{\rm 610MHz}$} & \colhead{Spectral Index} & Remarks \\ \cline{3-3}  
\colhead{source} &  & \colhead{$RA_{\rm J2000}~~~~~~~~DEC_{\rm J2000}$} &  & & \colhead{$\alpha$} & \\
 & \colhead{ID} & \colhead{(hh:mm:ss ~~~dd:mm:ss)} & \colhead{ (mJy)} &\colhead{ (mJy)} & ($S \propto \nu^\alpha$) &
}
\startdata
FGL\,J2018.1$+$4111 & S1 & 20:18:07.89 $+$41:10:41.6 & 7.5$\pm$1.0 &2.7$\pm$0.2 &  $-1.6\pm0.2$ &\\ 
3FGL\,J2018.6$+$4213 & S2 & 20:18:02.70 $+$42:06:52.6 & 28.5$\pm$1.0  & 15.5 $\pm$0.2 & $-1.0\pm0.1$ &\\
  & S3 * & 20:18:15.17 $+$42:17:32.8 & 5.8$\pm$0.8 &  4.8$\pm$0.3& $-0.3\pm0.3$ & double source\\
  & S4 * & 20:18:25.48 $+$42:11:31.7 & $<2.5$& 1.2$\pm$0.2 & $>-1.2$&\\
  & S5 * & 20:18:46.71 $+$42:20:04.6 & 400$\pm$100 & 245$\pm$5  & $-0.8\pm0.4$ & partially extended\\
  & S6 * & 20:19:01.99 $+$42:13:19.3 & 7.0$\pm$2.0 & 1.9$\pm$0.2 & $-2.1\pm0.5$&\\
  & S7 * & 20:19:17.81 $+$42:11:07.1 &  $<2.5$ & 1.3$\pm$0.2&   $>-1.0$&\\
  & S8 * & 20:19:18.53 $+$42:12:42.2 & 53$\pm$7 & 19.5$\pm$2.0 & $-1.6\pm0.3$ & overlapping w/ S9?\\
  & S9 * & 20:19:22.06 $+$42:12:32.4 & 78$\pm$6 & 22$\pm$2 & $-2.0\pm0.2$&  overlapping w/ S8? \\
4FGL\,J2021.5$+$4026 &  ---    &    &   [0.5] & [0.5]   & &\\
4FGL\,J2023.4$+$4127 & S10 & 20:23:35.71 $+$41:25:26.5 &  42$\pm$3 & 49.0$\pm$0.2 & $+0.2\pm0.1$ &\\ 
3FGL\,J2026.8$+$4003 &  ---    &    &   [0.6] & [1.0]   & &\\
3FGL\,J2028.5$+$4040c & S11 & 20:28:24.30 $+$40:37:49.3 & 347$\pm$9&  164$\pm$6&  $-1.2\pm0.1$&  double source\\
  & S12 * & 20:28:54.10 $+$40:44:07.8 & $<1.5$ & 1.1$\pm$0.5 & $>-0.5$&\\
4FGL\,J2030.9$+$4416 &  ---    &    &   [0.9] & [0.2]   & &\\
 4FGL\,J2032.2$+$4127 & S13 & 20:32:12.88 $+$41:27:24.2 &  2.8$\pm$0.2 & 0.8$\pm$0.1 & $-2.0\pm0.2$& \#5 of  Pa2007, PSR\\ 
4FGL\,J2032.6$+$4053 & S14 & 20:31:53.85 $+$40:55:19.1 & 5.5$\pm$0.3 & 2.0$\pm$0.2 & $-1.6\pm0.2$&\\   
  & S15 & 20:32:10.30 $+$40:52:58.2 & 2.7$\pm$0.4 & 2.8$\pm$0.3 & $+0.1\pm0.3$&\\
  & S16 & 20:32:25.72 $+$40:57:27.9 & 33$\pm$2 & 80$\pm$2 & $+1.4\pm0.1$&  Cyg\,X-3\\
  & S17 & 20:32:25.10 $+$40:50:15.7 & 4.7$\pm$0.4 & 2.0$\pm$0.2 & $-1.4\pm0.2$& \\
  & S18 & 20:32:34.22 $+$40:51:30.1 & 3.7$\pm$0.3 & 1.8$\pm$0.1 & $-1.1\pm0.2$&\\
  & S19 & 20:32:36.69 $+$40:55:58.2 & 11.0$\pm$0.4 & 5.1$\pm$0.2 &$-1.2\pm0.1$& AGN-a (SS2008)\\
  & S20 * & 20:32:38.89 $+$40:55:30.7 & 2.5$\pm$0.3 & 1.2$\pm$0.3 & $-1.2\pm0.4$& AGN-b (SS2008)\\
  & S21 & 20:32:39.77 $+$40:55:06.5 & 16$\pm$1 & 6.5$\pm$0.5 & $-1.4\pm0.2$&AGN-c (SS2008)\\
  & S22 & 20:32:38.56 $+$40:51:47.2 & 4.7$\pm$0.4 & 2.4$\pm$0.1 & $-1.1\pm0.2$& \\
  & S23 & 20:32:48.79 $+$40:48:05.5 & 11.0$\pm$0.6 & 14.0$\pm$0.2 & $+0.4\pm0.1$ &\\
  & S24 & 20:32:54.70 $+$40:54:49.3 & 36.0$\pm$1.5 & 22.5$\pm$0.4 & $-0.7\pm0.1$&\\
  & S25 & 20:33:07.30 $+$40:49:15.9 & 4.2$\pm$0.4 & 2.9$\pm$0.3 & $-0.5\pm0.2$ &\\
  & S26 * & 20:33:13.31 $+$40:50:56.4 & $<1.2$ & 1.6$\pm$0.2 & $>0.4$&\\
  & S27  & 20:33:14.80 $+$40:49:54.0 & 8.0$\pm$0.5 & 4.2$\pm$0.5 & $-1.0\pm0.2$& double source\\
3FGL\,J2032.5$+$4032 & S28 * & 20:32:08.40 $+$40:37:54.5 & 2.8$\pm$0.4  & 1.9$\pm$0.1  & $-0.6\pm0.2$&\\
   & S29 & 20:32:29.53 $+$40:38:50.2 & 40$\pm$1& 37.3$\pm$0.5 & $-0.1\pm0.1$&\\
   & S30 *& 20:32:34.12 $+$40:40:00.2 & 7$\pm$1 & 5.2$\pm$1.0& $-0.5\pm0.4$& partially extended\\  
   & S31 & 20:32:45.47 $+$40:39:38.2 & 28.6$\pm$0.5& 39.7$\pm$0.3& $+0.5\pm0.1$&\\
   & S32 & 20:32:52.33 $+$40:28:24.0  & 45$\pm$2 & 20$\pm$1 & $-1.3\pm0.1$& double source\\
   & S33 & 20:32:55.32 $+$40:31:31.5  & 225$\pm$4 & 127$\pm$3 & $-0.9\pm0.1$& double source \\
3FGL\,J2036.8$+$4234c  &  ---  &    &   [---] & [0.4]   & &\\
3FGL\,J2037.4$+$4132c & S34 & 20:37:26.61 $+$41:29:24.9 &  8.0$\pm$1.5&  5.0$\pm$1.5 & $-0.7\pm0.6$&\\
4FGL\,J2038.4$+$4212  & ---   &  &   [---]  & [0.6]   & &\\
3FGL\,J2039.4$+$4111  & S35 * &  20:38:56.99 $+$41:12:09.9 & 40$\pm$5  &  [---]  & &
\enddata
\tablecomments{The asterisks after the ID indicate radio sources not catalogued in \cite{benaglia2020b} in at least one band. Numbers in brackets are  rms values, and [---] means not observed. Upper limits in flux density correspond to 3$\sigma$ measured locally. AGNa,b,c (SS2008): components of an AGN discovered by \cite{sanchezsutil2008}. Pa2007: \cite{Paredes2007}. PSR = PSR\,J2032$+$4127 (pulsar).}
\end{deluxetable*}

For the gamma-ray sources within which no radio source was detected, we provide in Table\,\ref{tab:fluxs-spixs} the average rms at each observing band.

\subsection{Radio emission related to other gamma-ray sources}

\begin{figure*}[ht!]
\plotone{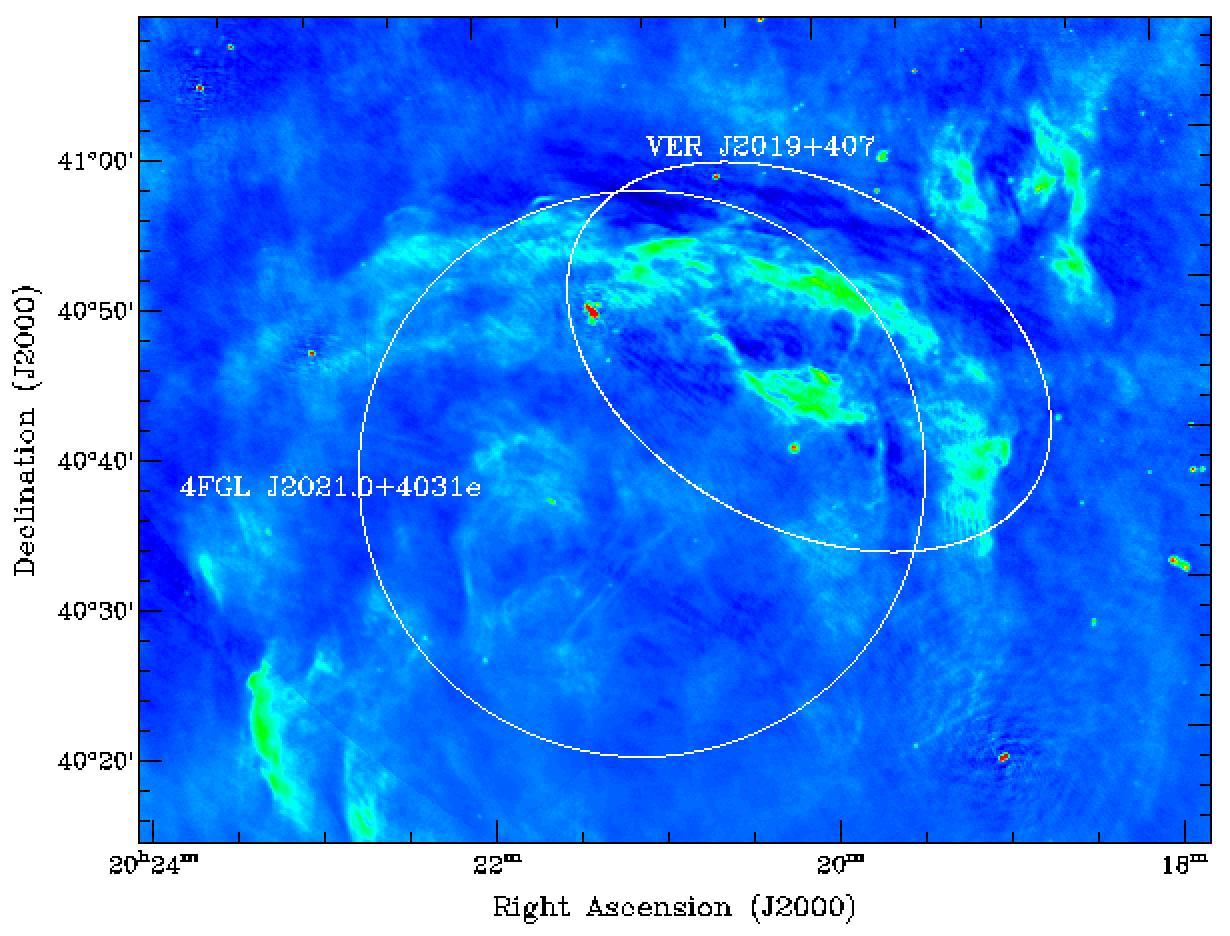}
\caption{325~MHz radio image of the SNR Gamma-Cygni (G78.2+2.1) with the overlaid contours of the Fermi source 4FGL\,J2021.0$+$4031e and the VERITAS source VER\,J2019$+$407. { The color scale interval shown is ($-$48,$+$83)~mJy~beam$^{-1}$, to outline weaker features}.}
\label{fig:ver2019-area}
\end{figure*}

\begin{figure}[ht!]
\centering
\includegraphics[width=8cm]{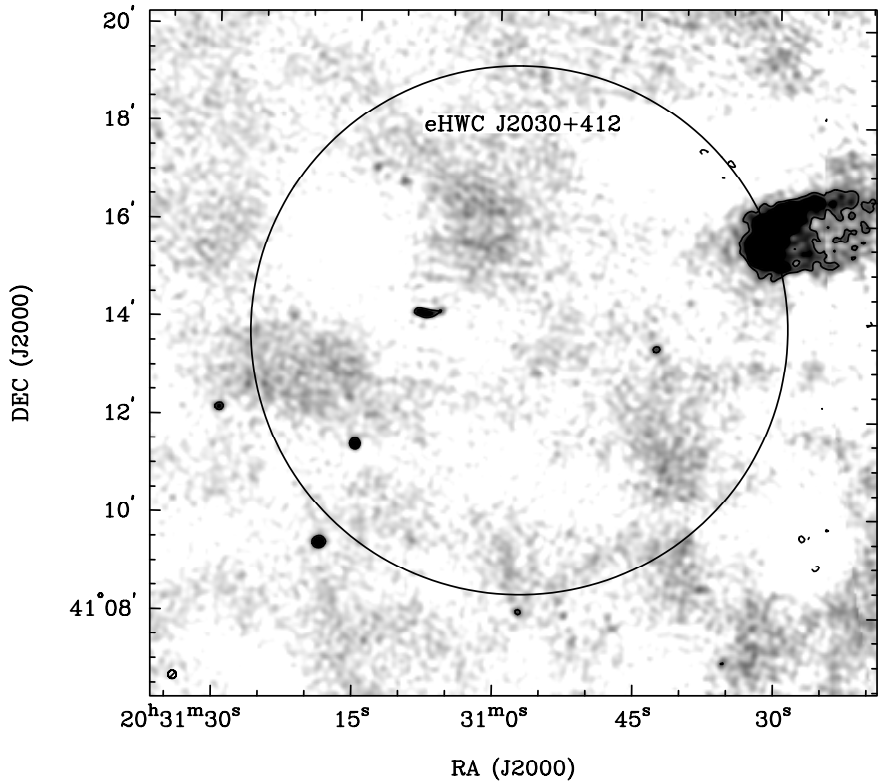}
\includegraphics[width=8cm]{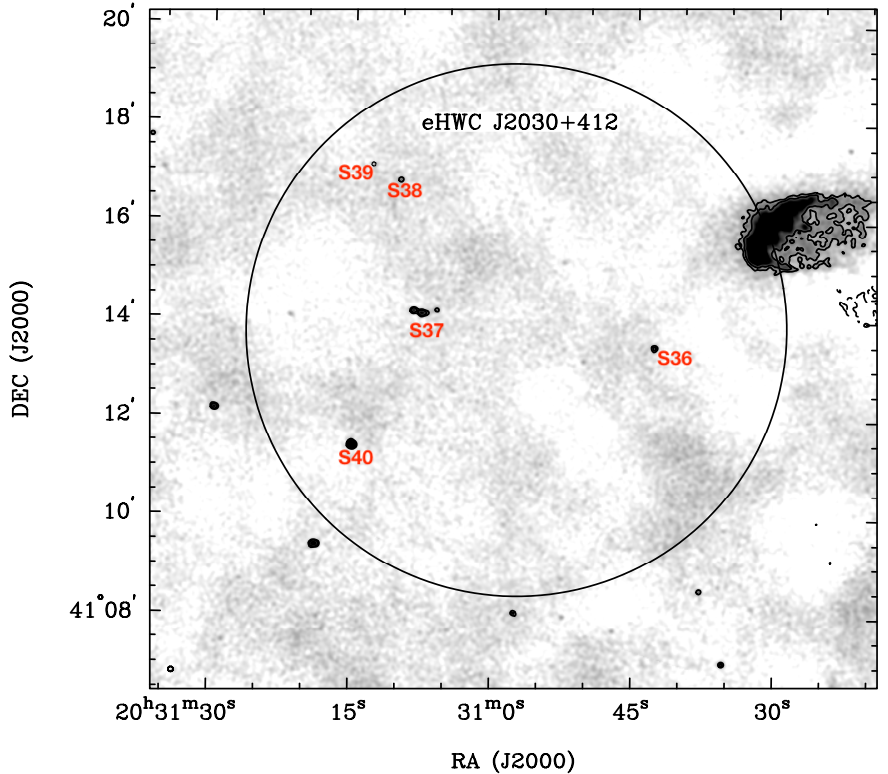}
\caption{eHWC\,J2030$+$412. Left: 325~MHz emission; contour levels $-3$, 3, 5, 10 and 20 in units of $\sigma$ (0.5~mJy beam$^{-1}$). Right: 610-MHz emission; contour levels $-3$, 3,  5, 10 and 20 in units of $\sigma$ (0.17~mJy beam$^{-1}$). The radio sources S36 to S40 are identified.}
\label{fig:ehwc}
\end{figure}

\begin{figure}[ht!]
\centering
\includegraphics[width=8cm]{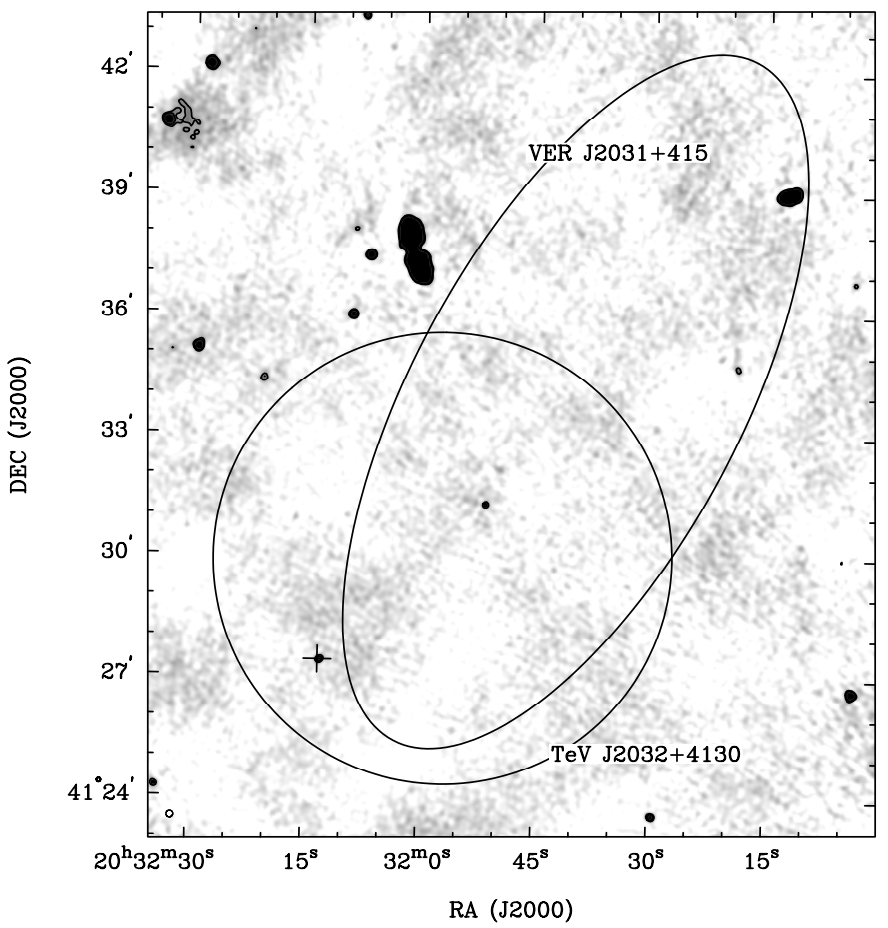}
\includegraphics[width=8cm]{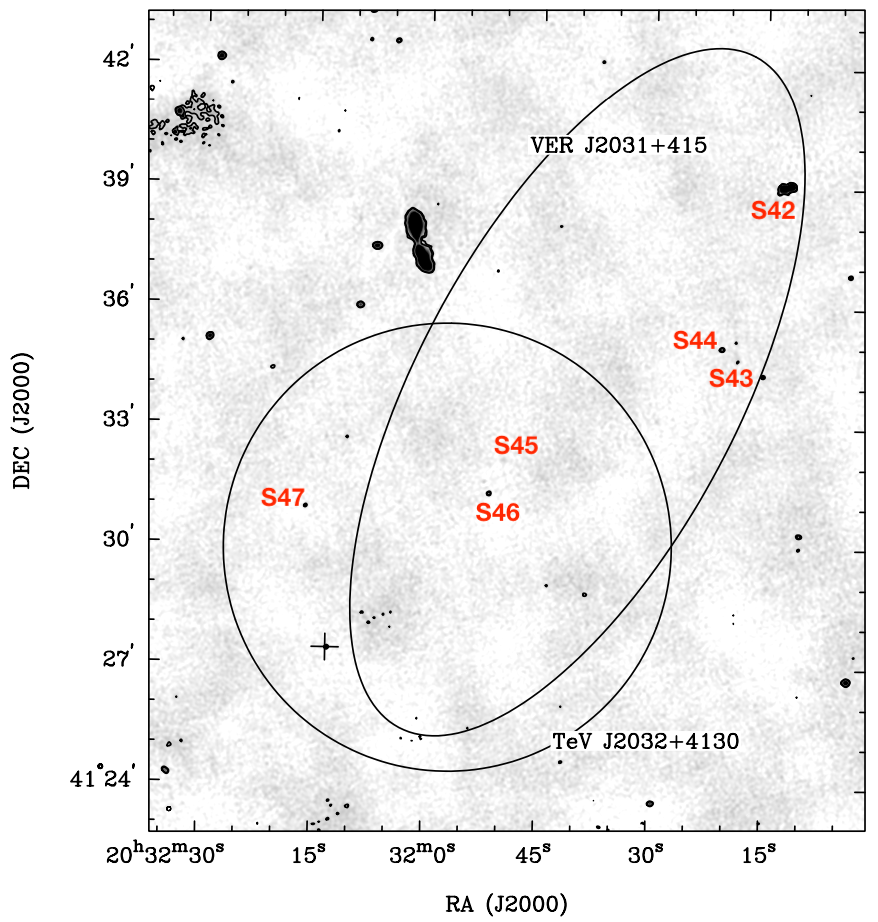}
\caption{TeV\,J2032$+$4130  and VER\,J2031$+$415. Left: 325~MHz emission; contour levels $-3$, 3 and 4.5 in units of $\sigma$ (0.4~mJy beam$^{-1}$). Right: 610-MHz emission; contour levels $-3$, 3 and 4.5 in units of $\sigma$ (0.12~mJy beam$^{-1}$). The radio sources S42 to S47 are identified.  {The cross is the pulsar PSR\,2032$+$4127 (and S13)}.}
\label{fig:tevver}
\end{figure}

\begin{deluxetable*}{@{}l l r r r r l}
\tablecaption{Flux density and spectral index of  radio sources detected in extended and/or TeV $\gamma$-ray sources}\label{tab:fluxs-spics2}
\tablewidth{0pt}
\tablehead{
\colhead{Gamma-ray} & \multicolumn{2}{c}{~~~~~~~~~~~Radio source} & \colhead{$S_{\rm 325MHz}$}  & \colhead{$S_{\rm 610MHz}$} & \colhead{Spectral Index} & Remarks \\ \cline{3-3}  
\colhead{source} &  & \colhead{$RA_{\rm J2000}~~~~~~~~DEC_{\rm J2000}$} &  & & \colhead{$\alpha$} & \\
 & \colhead{ID} & \colhead{(hh:mm:ss ~~~dd:mm:ss)} & \colhead{ (mJy)} &\colhead{ (mJy)} & ($S \propto \nu^\alpha$) &
}
\startdata
eHWC\,J2030+412 & S36 * & 20:30:42.92 $+$41:13:28.1 & 2.5$\pm$0.2 & 1.7$\pm$0.4 &  $-0.6\pm0.4$&\\ 
   &  S37 & 20:31:07.66 $+$41:14:07.3 & 7.0$\pm$0.3 & 5.4$\pm$0.3 &  $-0.4\pm0.1$& Ma2007, 
   triple source\\  
   &  S38 * & 20:31:10.16 $+$41:16:48.9 & 1.5$\pm$0.1 & 1.1$\pm$0.1 &  $-0.6\pm0.2$&\\
   &  S39 * & 20:31:12.93 $+$41:17:07.3  & 1.4$\pm$0.2  & 0.5$\pm$0.1 & $-1.6\pm0.4$ & Ma2007\\
   &  S40 & 20:31:14.93 $+$41:11:25.7 & 5.1$\pm$0.5 & 6.4$\pm$0.2 &  $+0.4\pm0.2$& Ma2007\\
4FGL\,J2028.6$+$4110e   & S41 * & 20:31:09.70 $+$42:30:24.6 &  5.9$\pm$1.0 & 1.9$\pm$0.2 &  $-1.8\pm0.3$&  \\ 
VER\,J2031$+$415  & S42  & 20:31:12.37 $+$41:39:03.9 & 84.7$\pm$2.0 & 40.0$\pm$2.0 &  $-1.2\pm0.1$&  Ma2007, double  source\\
  &  S43 * &  20:31:18.83 $+$41:34:43.9  & 1.9$\pm$0.3 & 0.6$\pm$0.2  & $-1.8\pm0.6$  &   Ma2007\\
  &  S44 * & 20:31:20.71 $+$41:34:59.3   & 1.0$\pm$0.2 & 1.0$\pm$0.2  & $+0.0\pm0.5$ &  Ma2007\\
TeV\,J2032+4130 &  S45 * & 20:31:49.69 $+$41:32:08.9  & 1.4$\pm$0.4 & 0.4$\pm$0.1 & $-2.0\pm0.6$ & \#2 of Pa2007\\
  &  S46 * &  20:31:51.59  $+$41:31:18.6  & 2.1$\pm$0.2 & 0.8$\pm$0.1 & $-1.5\pm0.3$ & \#3 of Pa2007\\
  & S47 * & 20:32:16.06 $+$41:30:56.0   & 0.8$\pm$0.2 & 0.6$\pm$0.1 & $-0.5\pm0.5$ &   \#6 of Pa2007\\
\enddata
\tablecomments{Asterisks as  in  Table\,\ref{tab:fluxs-spixs}. S45 and S46 correspond to the region where the  error ellipses of the two last VHE sources overlap. Ma2007: \cite{marti2007}, Pa2007: \cite{Paredes2007}. TeV\,J2032+4130 contains 4FGL\,J2032$+$4127, { and the radio source  S13} (see Table\,\ref{tab:fluxs-spixs} and also Pa2007).}
\end{deluxetable*}

{\bf 4FGL\,J2021.0$+$4031e} is an extended source associated with the supernova remnant Gamma-Cygni (SNR~G78.2$+$2.1).
Very  recently, the \cite{acciari2020} studied its energy spectrum and morphology by means of dedicated TeV observations  with the MAGIC Imaging Atmospheric Cherenkov telescopes.
VERITAS reported the discovery of an extended, unidentified source of VHE gamma rays, {\bf VER\,J2019$+$407}, lying along the northwestern shell of SNR~G78.2$+$2.1 \citep{Aliu2013}. 
Enhanced radio emission in this part of the SNR has also been observed at different frequencies, resolutions and sensitivities \citep[][and references therein]{Ladouceur2008}. Our results at 325~MHz also show clearly this enhanced emission but with a good sensitivity and resolution (Fig.\,\ref{fig:ver2019-area}), {despite for  4FGL\,J2021.0$+$4031e we found no discrete radio sources in a central region of 0.1~deg in size}.  A detailed study of the SNR~G78.2$+$2.1 is foreseen and deserves its publication in a future paper, { together with the radio sources in this extended field.}

We report five radio sources at both observing  bands inside the circle that represents {\bf eHWC\,J2030+412}. One of them, S37, is resolved at 610~MHz.
We derive the flux density values at both observing bands by measuring the emission above the 3$\sigma$ contour (see  Table\,\ref{tab:fluxs-spics2} and Fig.\,\ref{fig:ehwc}). 

{\bf 4FGL J2028.6+4110e} is a very extended gamma-ray source associated with the Cygnus Cocoon, MGRO J2031+41 and TeV J2032+4130. Due to its extension, we have not done a detailed study of the many sources in the area. A list of the sources detected at radio with more than 7$\sigma$ can be found in the catalog of \cite{benaglia2020b}. Here we only show {a  radio source at the very center  of  the  Fermi source}, S41, with very steep radio spectra (see Table\,\ref{tab:fluxs-spics2}), that  turns it a strong candidate for pulsar.  

{\bf TeV\,J2032$+$4130} was detected by HEGRA \citep{Aharonian2002, Aharonian2005}, and that was the first source discovered of a population of extended TeV sources without low-frequency counterparts. The discovery was later confirmed by MAGIC \citep{Albert2008} and VERITAS \citep{Aliu2014}, as  source {\bf VER\,J2031+415}. The pulsar PSR\,J2032$+$4127, detected by Fermi (4FGL\,J2032.2+4127), and S13 here, is within the field of TeV\,J2032$+$4130. The area had been observed with the  GMRT at 610~MHz by \cite{Paredes2007}. Apart from the detection of PSR\,J2032$+$4127 at 325 and 610~MHz, we have detected other sources (see Table\,\ref{tab:fluxs-spics2} and Fig.\,\ref{fig:tevver}) in the areas limited by the HEGRA, MAGIC and VERITAS;  one is  double  (S42). Maybe one of them could contribute additionally to the emission of this extended gamma-ray source because it is not clear if the pulsar alone can be responsible of the full emission. 

The search for radio counterparts related to sources of  Table\,\ref{tab:tevsources} provided twelve more candidates (S36 to  S47), presented in Table\,\ref{tab:fluxs-spics2}, totalling then 47 radio sources found { (see the figures in  the Appendix)}.

\section{Counterparts at other spectral ranges} \label{sec:counterparts}

In order to obtain additional information from the radio sources, we searched several large catalogs in the literature. These are listed in Table\,\ref{tab:catalogs} and offer results collected with instruments working from radio to X-ray bands. 
Regarding the search of radio counterparts on catalogs of massive, early-type stars, involving the same observations and images used here, see \cite{benaglia2020a}.

In the case of the 2MASS, WISE and Spitzer-SSCSL lists \citep{cutri2003,cutri2012,capak2013}, we found that a wealth of objects permeate the fields of the radio sources S1 to S47, and in the vast majority of the cases, their nature is not determined. Since the objects bright at infrared are usually thermal emitters, and we are interested here on non-thermal sources, we  considered the 2MASS, WISE and SSCSL sources that overlap the radio sources only in the cases for which the nature of the IR source was known. 

\begin{deluxetable*}{@{}l l l l}
\tablecaption{Catalogs used to look for counterparts to the radio sources S1 to S47\label{tab:catalogs}}
\tablewidth{0pt}
\tablehead{
\colhead{ID} & \colhead{Name} &  \colhead{Reference} & Associated symbol$^{\rm a}$
}
\startdata
C1 & TGSS Alternative Data Release  &   \cite{intema2017} & $25''$-side square\\  
C2 & The Westerbork Northern Sky Survey & \cite{rengelink1997} & $35''$-radius circle\\ 
C3 & Small-diameter radio sources catalogue & \cite{zoone1990} & $2'' \times  3''$ ellipse\\  
C4 & NRAO VLA Sky Survey & \cite{condon1998} & $25''$-radius circle\\   
C5 & The MIT-Green Bank 5GHz Survey   & \cite{bennett1986} & \\ 
C6 & 87GB Catalog of radio sources   &  \cite{gregory1991} & $3'' \times 2''$ ellipse\\ 
C7 & 2MASS All-Sky Catalog of Point Sources & \cite{cutri2003} & small  square\\ 
C8 & Spitzer Science Center Source List  & \cite{capak2013} & small cross\\ 
C9 & WISE All-Sky Data Release & \cite{cutri2012} & small circle \\  
C10 & 3XMM-Newton Serendipitous Source Catalogue-DR6  &  XMM-SSC, 2016 & small filled  circle\\ 
C11 & CHANDRA ACIS Survey of X-Ray Point Sources & \cite{wang2016} & small filled  circle\\ 
C12 & ATNF Pulsar Catalog  &  \cite{manchester2005} & \\ 
\enddata
\tablecomments{a: symbol used to represent the source in Figures\,\ref{fig:appe1}, \ref{fig:appe2}, \ref{fig:appe3}, and \ref{fig:appe4}.}
\end{deluxetable*}

\begin{deluxetable*}{l l l l l }
\tablecaption{Sources in catalogs  of Tables\,\ref{tab:fluxs-spixs} and \ref{tab:fluxs-spics2} positionally coincident with the radio sources S1 to S47\label{tab:counterparts}}
\tablewidth{0pt}
\tablehead{
\colhead{Source} & \colhead{Object name} & \colhead{Catalog$^{\rm a}$}  & \colhead{Type} & \colhead{Other reference}
}
\startdata
S5  & NVSS\,J201847+422010  & C1,C2,C4 & radio source &  \\ 
S10 & WISEA\,J202335.82+412523.5 & C9 & YSO &  \cite{kryukova2014} \\
S11 & NVSS\,J202824+403751 & C1,C2,C4 & radio source & \cite{wendker1991} \\
S13 & PSR\,J2032$+$4127 & C10,C11,C12 & pulsar &  \cite{abdo} \\
S16 & Cyg\,X-3 & C3,C4,C6,C10,C11 & HMXB & \cite{sanchezsutil2008} \\
S19 & 2XMM J203237.5+405557  & C10 &  X-ray source &    \\
S23 & CXOGSG J203248.8+404804 & C11 & X-ray source  & \\
S27 & 2MASS\,J20331489$+$4049096  & C7 & star &  \\
S29 & G079.602+0.506 & C3 & radio source & \cite{wendker1991} \\ 
S31 & MWC\,349 & C3,C4 & OB star(s) & \cite{cohen1985} \\
S33 & NVSS\,J203254+403128 & C2,C4 & radio source &  \cite{wendker1991} \\
S46 & XMM\,J203151.8$+$413118 & C10,C11 & X-ray  source &\\
\enddata
\tablecomments{a: We included sources of C7,  C8 and C9 only if their nature is known.}
\end{deluxetable*}

Table\,\ref{tab:counterparts} summarizes the results of the search for counterparts in the mentioned catalogs, to the radio sources S1 to S47. 
 
{ The individual  figures of  the radio sources S1 to S47 (Sect. Appendix)  show the positions of the counterparts found as well. The membership of the counterparts to a given catalog are shown with labels: C1 with ``TGSS'', C2 with ``WENSS'', C3 with ``1.4GHz'', C4 with  ``NVSS'', C6 with ``87GHz'', C10 with ``XMM'', and C11 with ``XCHA''. C7 sources are represented with small boxes, C8 sources with small crosses and C9 sources with small circles.}

We also compared the radio sources detected here with those listed in  
\cite{benaglia2020b}, a catalog of $\sim$3800 sources at the same Cygnus region carried out with the Giant Metrewave Radio Telescope at 325 and 610 MHz. The article provides the list of sources detected in the SPAM-generated mosaics (see Section 3), with  integrated flux densities above 7$\sigma$ (local rms). Radio sources with an  asterisk  in  Table\,\ref{tab:fluxs-spixs} are presented, at least at one band, here for the first time.

\section{Discussion}
 
\subsection{On the catalogs search results}

At the position of {\bf S5} lies the known sources  NVSS\,J201847$+$422010, WENSS\,B2017.0$+$4210 and TGSS\,J201846.6$+$422010. Their integrated flux densities, at 1400, 325, and 150~MHz, are $95.8\pm3.5$, $323$ and $620.6\pm84.6$~mJy (see  Fig.\,\ref{fig:appe1} and corresponding symbols according to  Table\,\ref{tab:catalogs}). The Gaussian fit to S5 with our GMRT data resulted in a discrete source, among errors. Regardless the different angular resolutions used in the various measurements collected, S5 clearly stands as a non-thermal source with a spectral index close to $-1$, if the flux values at 325~MHz are averaged. 

\cite{kryukova2014} reported a Young Stellar Object (J202335.81$+$412523.53) at the location of {\bf S10}. It showed a spectral index of $+0.2\pm0.1$. Although there are studies of such objects and  related types that predict, under certain conditions, gamma-ray emission from them, this is expected for those characterized by non-thermal spectral indices \citep[see, for instance,][]{bosch-ramon2010}. Thus, we do not expect high-energy contribution from this source  (Fig.\,\ref{fig:appe1}). 

{\bf S11} is another source with radio counterparts, as NVSS J202824+403751, WENSS B2026.6+4027 and TGSS J202824.3+403749; the integrated flux densities are $12.8\pm0.7$, $218$ and $392.8\pm44.2$~mJy, respectively  (see  Fig.\,\ref{fig:appe1}). \cite{wendker1991}
reported at the same position the source 19P06, with a flux at 1400~MHz of $64\pm3$~mJy, and \cite{taylor1996}, at  327~MHz,  WSRT\,2026$+$4027 with $126\pm7$~mJy. A mean spectral index  of $-0.8$ is obtained considering all measurements.

{\bf S13} is associated with the gamma-ray pulsar PSR\,J2032$+$4127. This source was first detected, with a flux density of 0.7$\pm$0.2 mJy, in a search of radio sources carried out by GMRT at 610 MHz in the field of the unidentified gamma-ray source TeV J2032+4130 \citep{Paredes2007}, measurement in full agreement with the flux reported here. XMM\,J203213.2$+$412723 (C10) and CHA\,J203213.1$+$412722 (C11) are other overlapping  sources  (see Fig.\,\ref{fig:appe2}).
The object, a binary system composed { by a pulsar} and a Be star, has been widely studied \citep[see for instance ][and references therein]{chernyakova2019}, and is listed in the 4FGL catalog as the counterpart to Fermi source 4FGL J2032.2$+$4127. S13 is the only radio source detected by us in the Fermi ellipse. {The spectral index derived here fits well for  what is expected for pulsars.}

\begin{figure}[ht!]
\centering
\includegraphics[width=8cm]{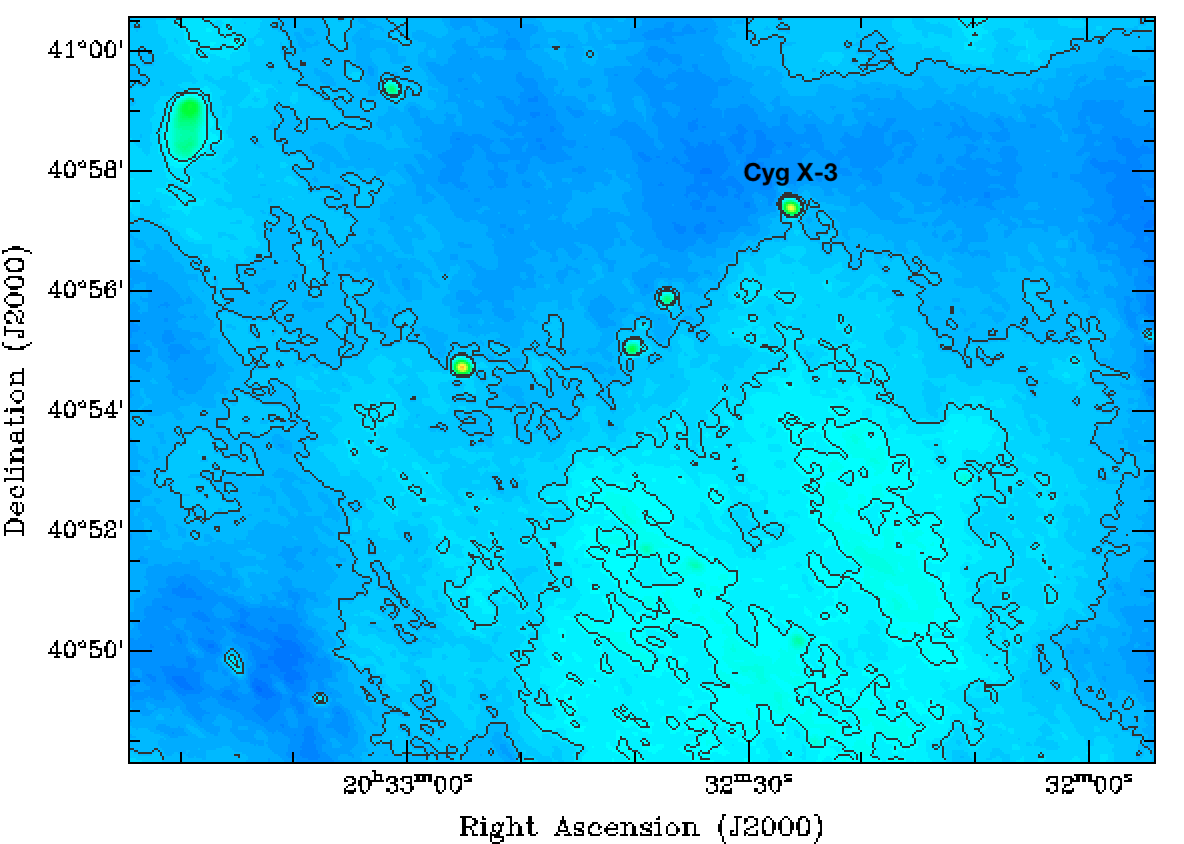}
\caption{Image at 325~MHz of the Cyg\,X-3 area, outlying diffuse exteded emission. {Color scale interval shown: ($-$51.3,$+$70.5)~mJy~beam$^{-1}$; contour levels: 0.3, 1.8 and 3.0 mJy~beam$^{-1}$}.}
\label{fig:cygx3-area}
\end{figure}

The source {\bf S16} agrees with the coordinates of the micro-quasar/High Mass X-ray Binary Cyg X-3, also a well known system, and associated with the source  4FGL\,J2032.6$+$4053. But contrary to the previous case, S16 is just one of the fourteen radio sources detected in the area of the Fermi source (see Fig.\,\ref{fig:appe2}). \cite{sanchezsutil2008} studied radio emission  around  this object at 2 and 6~cm, finding and extended diffuse radio structure south to Cyg\,X-3, and suggesting non-thermal emission. Our image at 325~MHz of the same area, built with a robust weighting of 0.0, showed such feature (see  Fig.\,\ref{fig:cygx3-area}). {At the position of S16, the catalog search yielded radio sources at 1.4~GHz (NVSS\,J203225$+$405728 of C4;  G079.846$+$0.700 of C3), and at 87~GHz (J203037.0$+$404726, C6), five C10  sources and one C11 source (symbols as per Table\,\ref{tab:catalogs}).}

Two X-ray sources are positionally coincident with {\bf S19} (XMM\, J203236.1$+$405603 and XMM\,J203237.5$+$405558).  Sources S19,  20 and  21 have  been identified  by \cite{sanchezsutil2008} as parts of an AGN. The spectral indices derived here confirm  that result.

{\bf S23} is another radio source with an X-ray counterpart, CXOGSG\,J203248.8$+$404804, detected by means of CHANDRA observations. 

A star identified as 2MASS\,J20331489$+$404804 is reported at the location  of {\bf S27}.  

The  radio  source {\bf S29} shows a counterpart detected with the VLA at 1.4 GHz,  G079.602$+$0.506 (C3), and also overlaps  the  source 19P18  \citep{wendker1991}.

{\bf S31} is identified with the source MWC\,349, proposed to be an  OB+OB system. It is also listed as a C4 source (NVSS\,J203245$+$403938). The radio spectrum  of this stellar system is analyzed in \cite{benaglia2020a} (see references therein).

At the position of {\bf S33} there are NVSS J203254+403128 (flux density of $23.1\pm0.9$~mJy),  and WENSS B2031.1+4021 (integrated flux of  102  mJy) sources, and \cite{wendker1991} reported the source 19P22. The global  spectral index remains negative.

In the case of {\bf S37}, we found a source at 610~MHz resolved in three components. Previous  work  on this area was carried out by \cite{marti2007} (see Fig.\,\ref{fig:appe4} of the Appendix). 
The authors discovered, with the GMRT at the same frequency band, two sources at the position  of  S37,    
with coordinates in agreement between error bars with the two brighter maxima  of S37. They reported flux densities of $0.95\pm0.11$~mJy, and $0.86\pm0.11$~mJy. We measure a slightly larger flux above the 3$\sigma$ contour corresponding to the position of the 2007 sources: 4.3$\pm$0.1~mJy, although two of the S27 maxima correspond to the  detection by \cite{marti2007}.

{\bf S39},  {\bf S40}, {\bf S42}, {\bf S43} and {\bf S44} were discovered by \cite{marti2007} at 610~MHz, 
with  flux densities of 0.53$\pm$0.11, 3.56$\pm$0.11, (12.06$+$15.51)$\pm$0.11, 0.57$\pm$0.11~mJy and 0.97$\pm$0.11, respectively. Except in the case of S40, their values for the point sources agree very well with those reported  here. 

The source S42 was also detected by \cite{taylor1996} as J2029$+$4129, with a flux density of 51$\pm$3~mJy (Fig.\,\ref{fig:appe4}).

{\bf S46} was detected at X-rays (XMM J203151.8+413118 of  C10, and CHA J203151.8+413119 of C11, Fig.\,\ref{fig:appe4}).

\cite{Paredes2007} reported the detection at 610~MHz of {\bf S45}, {\bf S46} and {\bf S47}, with flux densities of 0.50$\pm$0.16, 1.75$\pm$0.19~mJy and  0.77$\pm$0.18~mJy. The values for S45 and S47 are coincident with ours. For S46, the flux seems to have drop to a half in the decade between the two observations ($\sim$2006--2016) {and the spectral index is now more steeper ($-1.5\pm0.3$) than before ($-0.47\pm0.03$), although in both cases is non-thermal.}
The difference between the flux densities of S40 and S46 can be ascribed to some kind of variability, a common phenomenon in this type of MHz sources, and could explain the different spectral indices found. 

\subsection{Radio sources with steep spectral indices}

Among the 47 radio sources, there is a group that shows very negative $\alpha$ values. Such steep spectral indices are usually associated with pulsars. \cite{bates2013} performed a detailed study on the spectral distribution of this type of sources. They analyzed the spectral indices of $\sim$1300 pulsars at the radio range, and combined techniques of population synthesis and likelihood analyses to  find the spectral index distribution. They modeled the survey results with a Gaussian function in spectral index  distribution with the mean at $\alpha = -1.4\pm1$.

Later on, \cite{frail2016,frail2018} devised a method to pinpoint pulsar candidates, taking into account the spectral index and the compactness $C$ of radio sources at low radio frequencies. The  authors defined $C$ as the flux density over the peak brightness of the radio source. They studied TGSS-ADR1 together with NVSS sources, and showed that in the $\alpha\,C$ plane, the distribution for typical radio sources differed significantly from that of pulsars (see  their Figure\,1). The mean spectral index resulted in $\alpha = -0.73$ for the first group, and $\sim-1.8$ in the  case of pulsars \citep[see also ][]{intema2017}. Pulsar (candidates) share the extreme index values with luminous high-redshift {galaxies \citep{frail2016},  although it is not expected to receive VHE gamma-rays from the latter because absorption \citep{Biteau2015}}. 
\cite{frail2018} characterized promising pulsar candidates as having both $\alpha \leq -1.5$ and $C \leq 1.5$. The method was applied to TGSS-ADR1 and NVSS sources in the error ellipses of Fermi LAT UNIDs of 3FGL. Out of sixteen pulsar candidates identified, with follow-up continuum observations and timing, they found six ms pulsars and one normal pulsar. They considered that propagation effects through the ISM for instance, could prevent the pulsars to pop up during periodicity searches. The pulsed radio beams can even point off the line of sight \citep{camilo2012}.

Table\,\ref{tab:steeps} contains { eleven sources  with $\alpha \leq -1.5$, being one of them S13, the known pulsar PSR\,2032$+$4127 that is associated to 4FGL\,J2032.2+4127. Regarding their compact factor, in nine out of the eleven cases it remains lower or close to 1.5, another feature in common with known pulsars. And two radio sources, {S8 and S9,}  have very steep spectral indices with larger compact factors, and { the overall morphology resembles an AGN-like double source; all points} to an extragalactic origin.}

Leaving aside the resolved sources S8 and S9, the radio sources of Table\,\ref{tab:steeps} overlap infrared sources, that need to be studied as possible stars related to a pulsar. Their radio fluxes measured in this work are of the same order as those reported by \cite{frail2016}. None of them presented a 1.4~GHz counterpart (catalog C3). 

{ According to \cite{acero2015}, a Fermi source with variability index greater than 73 is considered as probably variable. The last column of Table\,\ref{tab:fermisources} shows that the variability indices of the Fermi sources for which the steepest spectral indices were measured, are below the value mentioned above. On the contrary, for the two Fermi sources showing values larger than 73, either we did not detect radio counterparts (4FGL\,J2021.5$+$4026) or the detected ones showed spectral indices $\geq-1.3$ (3FGL\,J2032.5$+$4032), thus no pulsar candidates.}

Finally, it is worth noticing that there are {seven} sources in Tables\,\ref{tab:fluxs-spixs} and \ref{tab:fluxs-spics2} showing spectral indices between $-1.5$ and $-1.2$. Three correspond to a previously proposed AGN \citep[S19, S20, S21, ][]{sanchezsutil2008}. Another three show flux densities between 20 and 350~mJy: S11, S32 and S42. Taking into  account the spectral index errors involved, the remaining, S17, might be also included in the pulsar candidates group.

\begin{deluxetable}{l l r r}
\tablecaption{Radio sources with very steep spectral index\label{tab:steeps}}
\tablehead{
\colhead{HE/VHE source} & \colhead{ID} & \colhead{$\alpha$} & $C$-factor$^{\rm a}$
}
\startdata
FGL\,J2018.1$+$4111  & S1  & $-1.6\pm0.2$ & 1.7\\
3FGL\,J2018.6$+$4213 & S6  & $-2.1\pm0.5$  & 1.1 \\
                     & S8  & $-1.6\pm0.3$ & 3.4 \\
                     & S9  & $-2.0\pm0.2$ & 4.9 \\
4FGL\,J2032.2$+$4127 & S13 & $-2.0\pm0.2$ & 1.0\\
4FGL\,J2032.6$+$4053 & S14  & $-1.6\pm0.2$ & 1.2\\
eHWC\,J2030$+$412    & S39  & $-1.6\pm0.4$ & 1.1\\
4FGL\,J2028.6$+$4110e & S41 & $-1.8\pm0.3$ & 1.0 \\
VER\,J2031$+$415 & S43 & $-1.8\pm0.6$ & 1.1 \\
TeV\,J2032$+$4130 & S45  & $-2.0\pm0.6$ & 1.5\\
  & S46 & $-1.5\pm0.3$ & 1.1\\
\enddata
\tablecomments{a: At the observed band at which is lower.  { The C factor of S8 and S9 is not lower than 1.5, which is a requirement  for pulsar candidate \citep{frail2016}.}}
\end{deluxetable}

\section{Conclusions}

We have performed wide area two frequency radio observations of the rich Cygnus 
region with the GMRT. There are several { HE/VHE} sources with counterparts not known in this region and we searched for radio sources in the error ellipses of them to look for counterparts.
We found several interesting radio sources within the error circle of the HE/VHE sources. The main conclusions are the following:

\begin{itemize}
\item In the error ellipses of 8 HE/VHE sources we detected 11 radio sources with very steep spectral index ($\alpha \leq -1.5$). Nine of them have a C-factor $\lesssim 1.5$, indicating that are pulsar candidates.

\item Two sources, S40 and S46, show variability when compared to previous GMRT data at 610 MHz. Other sources (8), of which  previous arcsec-resolution data were available, are steady whereas in 37 cases there is no previous data to compare.

\item We have obtained a 325 MHz image of the supernova remnant Gamma-Cygni (G78.2+2.1) which contain 4FGL\,J2021.0+4031e and VER\,J2019+407. In a central region of
0.1 deg in size of  4FGLJ2021.0+4031e we
found no discrete radio sources.

\item No radio sources were detected above 2~mJy level for 4FGL J2021.5$+$4026, 3FGL J2026.8$+$4003, 4FGL J2030.9$+$4416, 3FGL J2036.8$+$4234c and 4FGL J2038.4+4212. A radio source at 325~MHz was detected in the error ellipse of  3FGL\,J2039.4$+$4111.\\
\end{itemize}

The detection  of radio sources with characteristics of pulsar candidates prompts for further investigation to confirm their nature, and from then on, to describe the extent to which they can contribute to the (very) high-energy sources.
A follow-up study on timing of such sources with the Very Large Array, GMRT or VLBI facilities is needed to verify its nature. As quote \cite{frail2016}, such very high angular resolution is needed to discriminate resolved luminous high-redshift  galaxies from unresolved pulsars.

\acknowledgments

We thank the staff of the GMRT that made these observations possible. GMRT is run by the National Centre for Radio Astrophysics of the Tata Institute of Fundamental Research. We are also grateful to Josep Mart\'{\i} for useful comments.
PB acknowledges support from ANPCyT PICT 0773--2017. 
CHIC acknowledges the support of the Department of Atomic Energy, Government of India, under the project  12-R\&D-TFR-5.02-0700. 
JMP acknowledges financial support by the Spanish Ministerio de Economía, Industria y Competitividad (MINEICO/FEDER, UE) under grant AYA2016-76012-C3-1-P, from the State Agency for Research of the Spanish Ministry of Science and Innovation under grant PID2019-105510GB-C31 and  through the ``Unit of Excellence María de Maeztu 2020-2023'' award to the Institute of Cosmos Sciences (CEX2019-000918-M), and by the Catalan DEC grant 2017 SGR 643.
This research has made use of the SIMBAD and Vizier databases, operated at CDS, Strasbourg, France, and of NASA's Astrophysics Data System bibliographic services.

\appendix\label{appen}

\section*{Figures of the radio sources S1 to S47}

The ID of the radio source, the imaged band, and the contour levels with its unit are shown in each sub-figure. 

\begin{figure*}
\centering
\includegraphics[width=17cm]{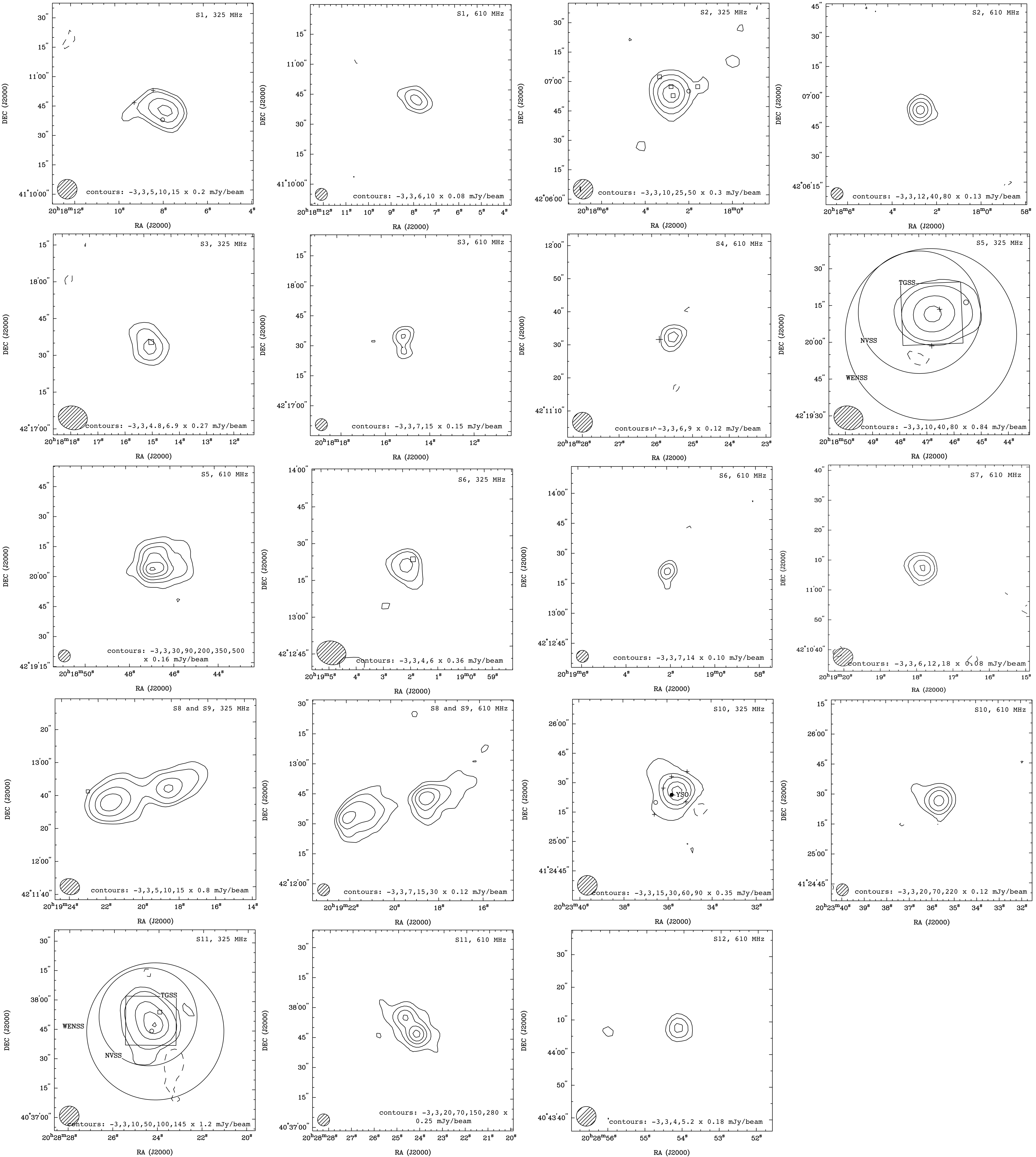}
\caption{Figures of radio sources S1 to S12, at both observed frequency bands, except  for S4, S7 and S12 which were only  detected  at 610~MHz. Counterparts found in the literature are also shown; the different symbols used are listed in Table 5, last column. For S12, the position of the YSO reported by \cite{kryukova2014} is marked.}
\label{fig:appe1}
\end{figure*}

\begin{figure*}
\centering
\includegraphics[width=17cm]{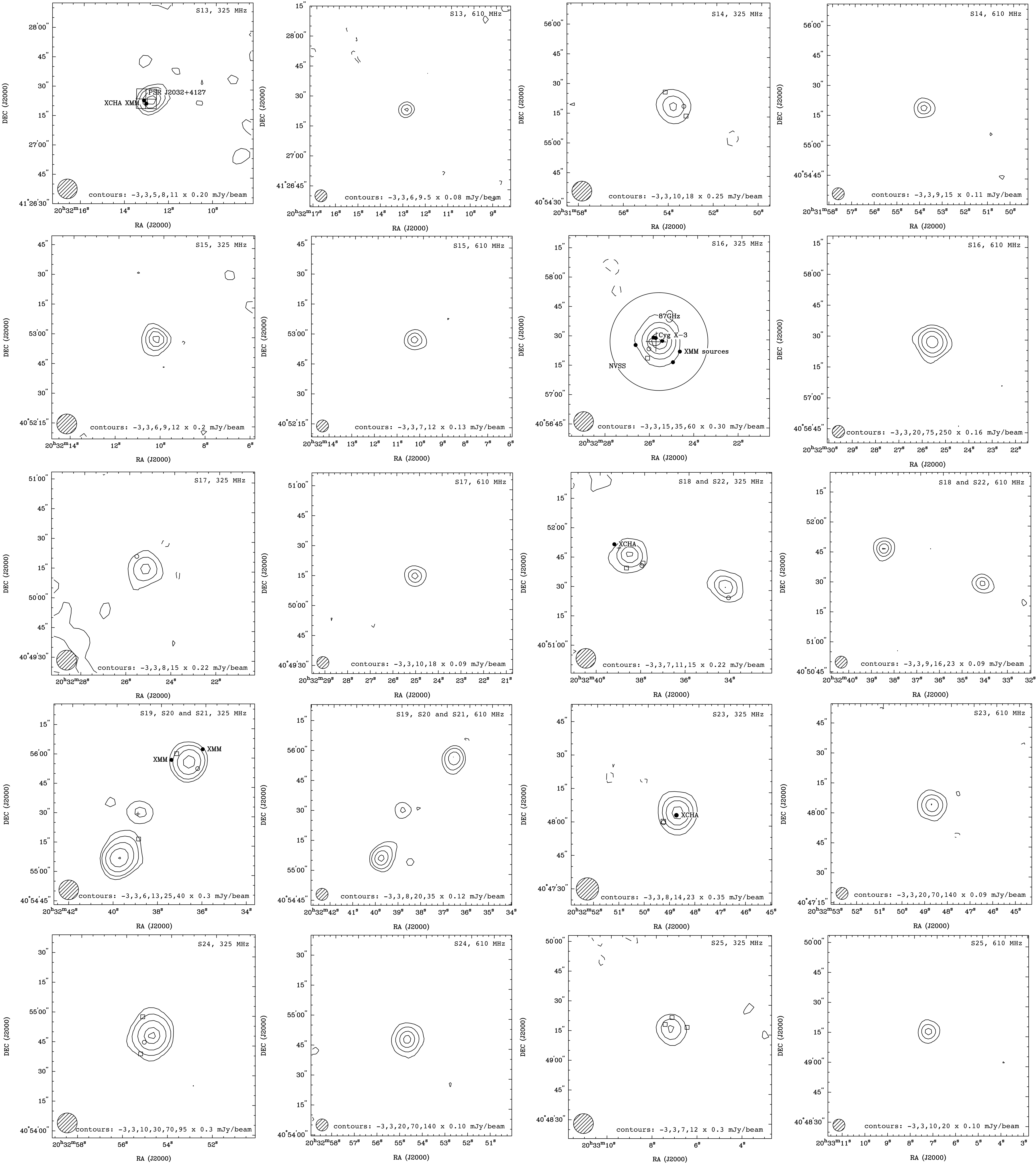}
\caption{Figures of radio sources S13 to S25, at both observed frequency bands. Counterparts found in the literature are also shown; the different symbols used are listed in Table 5, last column. For S13, the box represents the source detected by \cite{Paredes2007}, and the  big  cross that of the pulsar  PSR\,J2032$+$4127. For  S16, the big cross marks the position of Cyg\,X-3.}
\label{fig:appe2}
\end{figure*}

\begin{figure*}
\centering
\includegraphics[width=17cm]{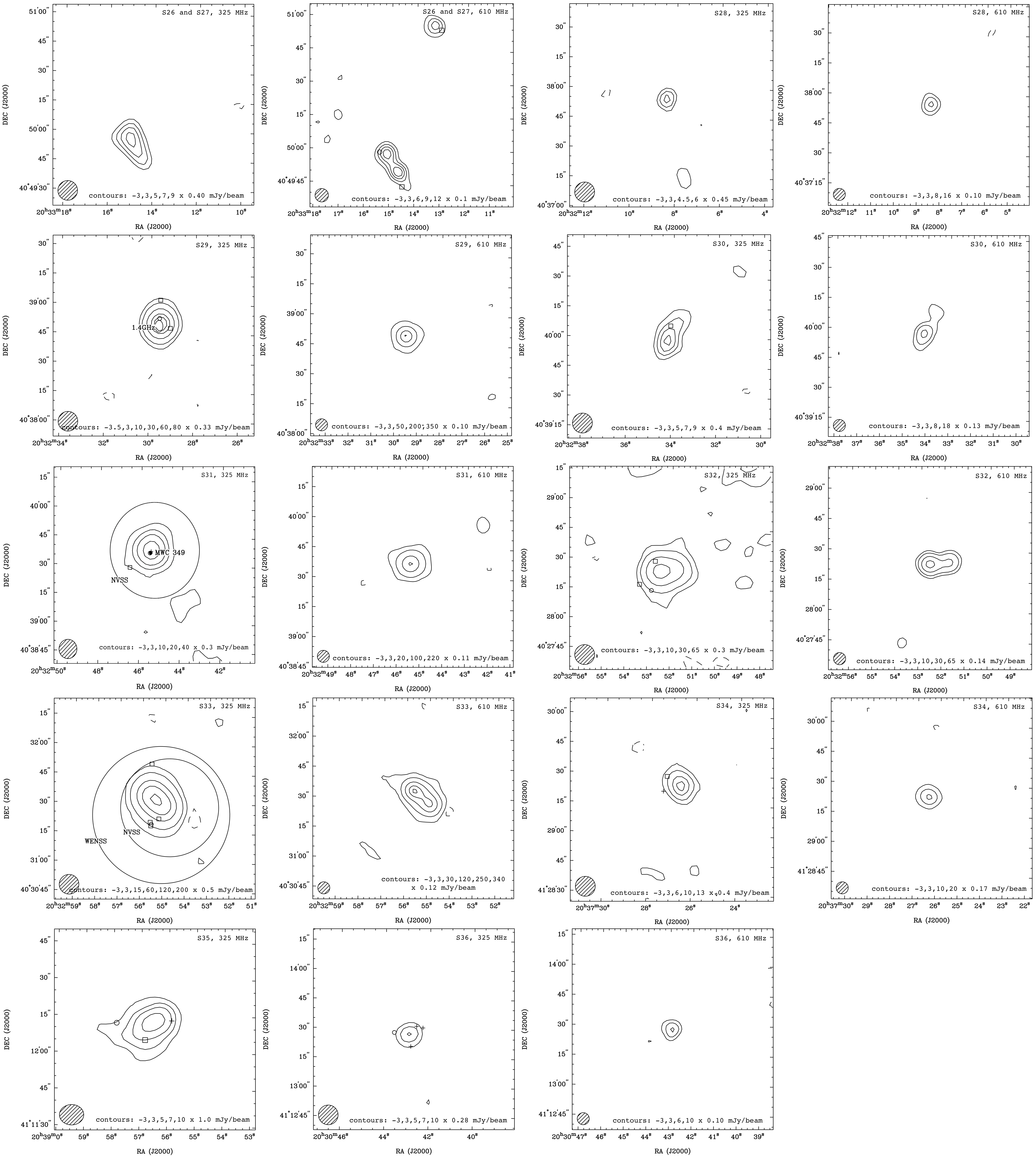}
\caption{Figures of radio sources S26 to S36, at both observed frequency bands, except for S35 which was only observed  at 325~MHz. Counterparts found in the literature are also shown; the different symbols used are listed in Table 5, last column.}
\label{fig:appe3}
\end{figure*}

\begin{figure*}
\centering
\includegraphics[width=17cm]{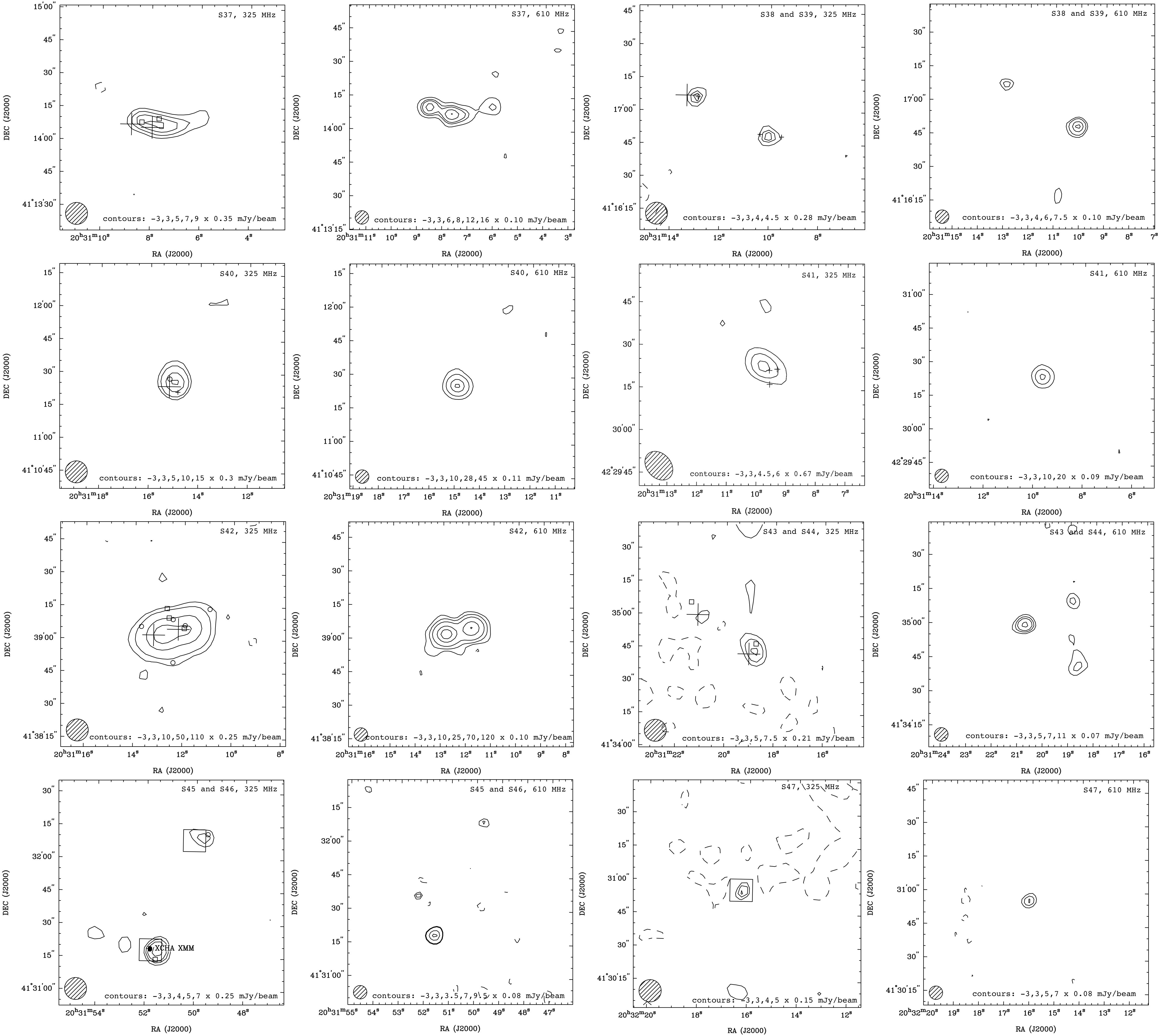}
\caption{Figures of radio sources S37 to S47, at both observed frequency bands.  Counterparts found in the literature are also shown; the different symbols used are listed in Table 5, last column. For S37, S39, S40, S42, S43 and S44, the  big crosses mark the sources reported by \cite{marti2007}. For S45, S46 and S47, the boxes represents the source detected by \cite{Paredes2007}. }
\label{fig:appe4}
\end{figure*}

\bibliography{HEsources}{}
\bibliographystyle{aasjournal}
 
\end{document}